\documentclass[iop,revtex4-1]{emulateapj}

\usepackage{graphicx}
\usepackage{amssymb}
\usepackage{amsmath}

\usepackage{natbib}
\usepackage[backref,breaklinks,colorlinks,citecolor=blue]{hyperref}
\usepackage{subfigure}
\usepackage[flushleft]{threeparttable}
\usepackage{etoolbox}

\usepackage{color}

\makeatletter

\patchcmd{\NAT@citex}
  {\@citea\NAT@hyper@{%
     \NAT@nmfmt{\NAT@nm}%
     \hyper@natlinkbreak{\NAT@aysep\NAT@spacechar}{\@citeb\@extra@b@citeb}%
     \NAT@date}}
  {\@citea\NAT@nmfmt{\NAT@nm}%
   \NAT@aysep\NAT@spacechar\NAT@hyper@{\NAT@date}}{}{}

\patchcmd{\NAT@citex}
  {\@citea\NAT@hyper@{%
     \NAT@nmfmt{\NAT@nm}%
     \hyper@natlinkbreak{\NAT@spacechar\NAT@@open\if*#1*\else#1\NAT@spacechar\fi}%
       {\@citeb\@extra@b@citeb}%
     \NAT@date}}
  {\@citea\NAT@nmfmt{\NAT@nm}%
   \NAT@spacechar\NAT@@open\if*#1*\else#1\NAT@spacechar\fi\NAT@hyper@{\NAT@date}}
  {}{}

\makeatother

\shorttitle{The formation of a NSC around a MBH in the Henize 2-10 galaxy}
\shortauthors{Arca-Sedda, Capuzzo-Dolcetta, Antonini, Seth}

\begin{document}

\title{Henize 2-10: the ongoing formation of a nuclear star cluster around a massive black hole}
\author{M. Arca-Sedda\altaffilmark{1,2}}
\author{R. Capuzzo-Dolcetta\altaffilmark{2}}
\affil{Dept. of Physics, Universit\`{a} di Tor Vergata, Via O. Raimondo 18, I-00173 Rome, Italy}
\affil{Dept. of Physics, Sapienza-Universit\`{a} di Roma, P.le A. Moro 5, I-00165 Rome, Italy}
\author{F. Antonini\altaffilmark{3}}
\affil{Center for Interdisciplinary Exploration and Research in Astrophysics (CIERA) and Department of Physics and Astronomy, Northwestern University, 2145 Sheridan Road, Evanston, IL 60208, U.S.A.}
\and 
\author{A. Seth\altaffilmark{4}}
\affil{Physics and Astronomy Dept, University of Salt Lake city, 201 Presidents Cir, Salt Lake City, UT 84112, USA}

\email{m.arcasedda@gmail.com}

\begin{abstract}
The central region of the galaxy Henize 2-10 hosts a black hole (BH) candidate with a mass $Log\left(M_{\rm BH}/{\rm M}_{\odot}\right)=6.3\pm 1.1$. While this putative black hole does not appear to coincide with any central stellar overdensity, it is surrounded by 11 young massive clusters with masses above $10^5$ M$_\odot$. The availability of high quality data on the structure of the galaxy and the age and mass of the clusters provides excellent initial conditions for studying the dynamical evolution of Henize 2-10's nucleus. Here we present a set of $N$-body simulations in which we model the future evolution of the central clusters and the black hole to understand whether and how they will merge to form a nuclear star cluster. Nuclear star clusters (NSCs) are present in a majority of galaxies with stellar mass similar to Henize 2-10. While the results depend on the choice of initial conditions, we find that a NSC with mass $M_{\rm NSC}\simeq 4-6\times 10^6$ M$_\odot$ and effective radius $r_{\rm NSC}\simeq 2.6-4.1$ pc will form within $0.2$ Gyr. This work is the first showing, in a realistic realization of the host galaxy and its star cluster system, that the formation of a bright nucleus is a process that can happen after the formation of a central massive BH leading to a composite NSC+BH central system. The cluster merging process does not significantly affect the kinematics of the BH; when a stationary state is reached its position changes by $\lesssim 1$~pc and its velocity by $<2$ km s$^{-1}$.
\end{abstract}


\section{Introduction}
The majority of the observed galaxies with luminosities up to $10^{11}$L$_\odot$ host bright stellar nuclei usually referred to as Nuclear Star Clusters (NSCs) or ``resolved stellar nuclei''.
Such systems are observed in galaxies along the whole Hubble sequence \citep{rich,BKR02,cote06,Turetal12,denB} and they are the densest stellar systems known in the Universe, due to their high masses (up to $10^8$M$_\odot$) and small scale radii (only few pc). Moreover, it is not rare to find in galactic nuclei both a NSC and a super massive black hole (SMBH) \citep{seth08,graham09}. In particular, 
the center of galaxies with masses of $\sim 10^9$M$_\odot$ are typically dominated by the presence of a NSC, while a MBH could be present but it is likely unseen due to resolution limit of present instruments. On the other hand, galaxies with masses above $10^{11}$M$_\odot$ are dominated by the presence of SMBHs. Between such ranges, instead, SMBHs and NSCs have similar masses \citep{graham09,Neum,kormendy13}. 
\\
The fact that NSCs seem to dominate the center of smaller galaxies while SMBHs dominate heavier galactic centers suggests a link between these objects which, however, remains poorly understood.
\\
In the last years, much research has been devoted to examining scaling relations connecting the SMBHs or NSCs and their hosts. For instance, \cite{frrs} and \cite{rossa} have shown that the relation between the NSC mass and the host velocity dispersion is similar to that provided for SMBHs. On the other hand, more recent studies have shown that this relation is shallower for NSCs \citep{LGH,ERWGD,scot}.  
The correlation between the central object (SMBH or NSC) and the host could provide important information about the central object's formation history and evolution.
\\
At present there are several formation models for NSCs. In the scenario commonly referred to as ``in-situ'' formation \citep{King03,King05,Mil04,Beketal06,aharon15}, the NSC forms if the time-scale on which the BH can grow by accretion is larger than the crossing time of the galaxy \citep{nayakshin,hopkins10}. Such a scenario provides a relation between the NSC mass and the host velocity dispersion similar to that observed for SMBHs.
Other works proposed that a NSC forms first in the galactic nucleus,  and then its high density facilitate runaway collisions of massive stars, leading to the formation of a SMBH seed, which may subsequently grow through stellar and gas accretion \citep{zwart04}.
\\
On the other hand, NSCs can form via the well-studied
``dry-merger'' scenario, in which star clusters orbitally decay toward the galactic center under the action of the dynamical friction process \citep{TrOsSp,Trem76,Dolc93,CDMB09}. 
The subsequent merging of the decayed star clusters leads to the formation of a dense stellar nucleus with characteristics comparable to those of real nuclei.
In this scenario, the presence of a central BH in the host galaxy could lead to the disruption of the decaying clusters, preventing NSC formation. However, \cite{antonini13} and \cite{ASCD14b} have shown that the tidal forces induced by the central SMBH are only significant at BH masses above $10^8$M$_\odot$.
\\
The dry-merger scenario has been investigated through numerical simulations \citep{DoMioA,DoMioB,AMB,perets14}, and also via statistical and theoretical models \citep{LGH,antonini13,ASCD14b}. 
These works have shown that such a scenario could explain the observed properties of the nuclei, and also provide an explanation for the observed NSC-host galaxy scaling relations recently updated by several works \citep{LGH,ERWGD,scot}.
\\
In this context, galaxies with star clusters near their photometric center provide a perfect laboratory to test the dry-merger scenario. Two examples of such galaxies are the Fornax dSph galaxy, the heaviest satellite galaxy of the Milky Way \citep{buonanno,buonanno98,mackey,dinescu,coleman05,goerdt06,readcole,cole}, and the Henize 2-10 dwarf starburst galaxy \citep{kobul95,johnson2000,johnson,Otta,Ottb,santangelo,reines12}. A recent paper provided detailed constraints on the galaxy and star cluster properties of Henize 2-10 \citep{ngu14}. This galaxy likely hosts an accreting SMBH \citep{merloni,kobul} of mass ${\rm Log}(M/M_\odot)=6.3\pm 1.1$ \citep{reines} and eleven young super star clusters (SSCs) with masses above $10^5$ M$_\odot$, placed at projected distances $\lesssim 140$ pc from the galactic center \citep{ngu14}.  On the other hand, no NSC has been detected. The detailed informations available on the projected mass profile of the galaxy and the cluster masses and sizes, make Henize 2-10 an interesting case study for testing the dry-merger scenario.
\\
In this paper we  make use of direct-summation $N$-body simulations to investigate whether the future evolution of the young star cluster system observed in Henize 2-10 could lead to the formation of a central stellar nucleus, distributing the SSCs with initial conditions consistent with observational constraints on their orbit and internal properties.
\\
The paper is organized as follows: in Section~\ref{models} we introduce the models used to represent galaxy and SSCs and give an estimate of the mass growth of the expected NSC by using proper semi-analytical arguments; in Section~\ref{nbody} the results of the simulations are presented and discussed; in Section \ref{long} we discuss the possible long-term evolution of the NSC in Henize 2-10; finally, Section~\ref{end} is devoted to a final discussion and conclusions.

\section{Modeling the galaxy and the globular cluster system}
\label{models}
\subsection{The galaxy model}
\label{galmod}

 Data provided by \cite{ngu14} give virial estimates of the mass enclosed within two different values of the distance to the center of Henize 2-10:
\\
\begin{eqnarray}
M(r<2")  &=& (2.7\pm 1.1 )\times 10^8 \rm{M}_\odot , \\
M(r<6") &=& (6.4 \pm 1.5 )\times 10^8 \rm{M}_\odot ;
\label{hmass}
\end{eqnarray}
\\
as well as the effective radius of the inner component of the galaxy, $R_e=6"$. At the estimated distance of $9.0$ Mpc to us \citep{vacca92}, the above values refer to the mass enclosed within $89$ and $259$ pc, respectively.
\\
We modeled the galaxy  using the density profile model:
\begin{equation} 
\rho_\gamma(r)=\frac{(3-\gamma)M_H}{4\pi r_s^3}\left(\frac{r}{r_s}\right)^{-\gamma}\left(\frac{r}{r_s}+1\right)^{\gamma-4}\frac{1}{{\mathrm{cosh}(r/r_{\rm tr})}},
\label{den}
\end{equation}
where  $M_H$ is the total mass of the galaxy, $r_s$ its scale radius,  $\gamma$ is the inner density profile slope,
and $r_{\rm tr}$ is the truncation radius of the model. 
At radii $r \ll r_{\rm tr}$
the model corresponds to a Dehnen density profile \citep{Deh93} with mass distribution:
\begin{equation}
M_\gamma(r)=M_{\rm H}\left(\frac{r}{r+r_s}\right)^{3-\gamma}.
\label{dehnmass}
\end{equation}
The scale length, $r_s$, is related to  the observed effective radius, $R_e$,  through the relation
\begin{equation}
r_s = \frac{4}{3}(2^{1/(3-\gamma)}-1)R_e.
\end{equation}
\\

\begin{table}
\caption{}
\centering{Adopted parameters of the super-star clusters.}
\begin{center}
\begin{tabular}{crrrrr}
\hline
\hline
\multicolumn{1}{c}{ID} & \multicolumn{1}{c}{$M_{\rm SSC}$} & \multicolumn{1}{c}{$r_c$} & \multicolumn{1}{c}{$r_{\rm eff}$} & \multicolumn{1}{c}{$R_{\rm SSC}$} & \multicolumn{1}{c}{$N$}\\
& \multicolumn{1}{c}{$(10^6$M$_\odot)$} & \multicolumn{1}{c}{(${\rm pc}$)} & \multicolumn{1}{c}{(${\rm pc}$)} & \multicolumn{1}{c}{(${\rm pc}$)} & \\ 
\hline 
C1  & $2.30 \pm 0.60$ & $1.27 \pm 0.03$ & $3.1 $ & $64.8 \pm 1.4$  &$28976$ \\
C2  & $0.92 \pm 0.61$ & $0.77 \pm 0.01$ & $1.9 $ & $36.9 \pm 1.4$  &$11590$ \\
C3  & $1.14 \pm 0.78$ & $1.00 \pm 0.02$ & $2.4 $ & $20.4 \pm 1.4$  &$14362$ \\
C4  & $0.91 \pm 0.60$ & $0.71 \pm 0.05$ & $4.2 $ & $11.2 \pm 1.4$  &$11464$ \\
C5  & $0.40 \pm 0.21$ & $0.53 \pm 0.07$ & $1.3 $ & $86.4 \pm 1.4$  &$5039$ \\
C6  & $0.40 \pm 0.21$ & $0.53 \pm 0.04$ & $1.3 $ & $77.6 \pm 1.4$  &$5039$ \\
C7  & $0.46 \pm 0.25$ & $0.89 \pm 0.21$ & $2.2 $ & $114.5 \pm 1.4$ &$5795$ \\
C8  & $0.45 \pm 0.24$ & $0.46 \pm 0.02$ & $1.1 $ & $130.7 \pm 1.4$ &$5669$ \\
C9  & $0.20 \pm 0.11$ & $0.37 \pm 0.01$ & $0.9 $ & $74.1 \pm 1.4$  &$2519$ \\
C10 & $0.45 \pm 0.24$ & $0.46 \pm 0.02$ & $1.1 $ & $70.2 \pm 1.4$  &$5669$ \\
C11 & $0.20 \pm 0.11$ & $0.37 \pm 0.01$ & $0.9 $ & $55.8 \pm 1.4$  &$2519$ \\
\hline
\end{tabular}
\end{center}
\begin{tablenotes}
	\item Column 1: cluster name. Column 2: mass. Column 3: core radius. Column 4: effective radius. Column 5: projected radial coordinate. Column 6: number of particles used.
\end{tablenotes}
\label{tab1}
\end{table}

\begin{figure}
\centering
\includegraphics[width=8cm]{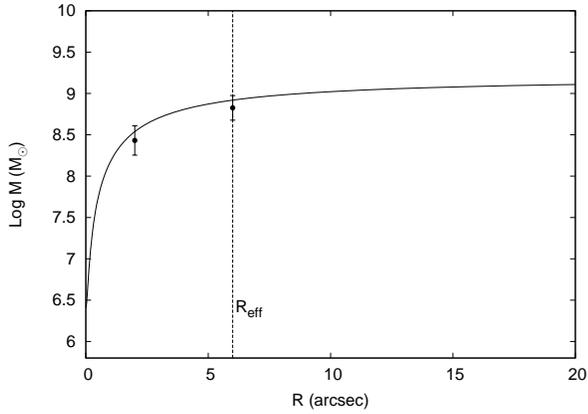}
\caption{Cumulative projected mass distribution (solid line) as obtained by Equation \ref{den} with the values given in Subsect.~\ref{galmod}, compared with the measured values of the mass enclosed within $2"$ ($86$ pc) and $6"$ ($259$ pc). The vertical dashed line represents the effective radius of the galaxy.}
\label{masspro}
\end{figure}

To limit the number of particles in our direct $N$-body simulations of Henize 2-10 we decided to restrict our self-consistent, particle, representation of the galaxy to the inner radial region. We show that this simplification provides a good description of the environment in which the clusters move and discuss the resolution of the simulations in Section~\ref{numbprob}. 
\\
Due to this choice, in what follows we set $M_{\rm H}=1.6\times 10^9$ M$_\odot$, $\gamma=1/2$ and $r_{\rm tr}=150$ pc. 
This choice accurately represents
the deprojected brightness profile of the galaxy 
at $r \lesssim 300$ pc and leads to a value of the galaxy mass within
$6"$ in good agreement with Equation \ref{hmass}. Moreover, the projected velocity dispersion in the range $1"-6"$ has a value of $41$ km s$^{-1}$, very close to the observed value \citep{marquart,ngu14}, ensuring a reliable representation of the dynamics of the clusters within the galaxy.
 It is worth noting that the high resolution of the data provided in \cite{ngu14}, $\sim 0.1"$ , implies that an error on the positional center of the galaxy does not affect significantly our choice of the galaxy model. 

In Figure \ref{masspro} we compare the mass profile of our model with the  mass profile obtained from the photometric data of \cite{ngu14}, assuming a distance of $d=9.0$ Mpc \citep{vacca92}.

Since there is observational evidence that the galaxy hosts a MBH at its center with mass ${\rm Log} (M_{\rm BH}/{\rm M}_\odot) = 6.3\pm 1.1$, we decided to include it in our models, generating the $N$-body representation of the whole galaxy via a numerically calculated distribution function.
While the mass of the BH candidate is quite uncertain, and therefore it would be worth testing models with different BH masses, the numerical simulations presented here are extremely time consuming. Therefore, we examine only one value for the BH mass ($M_{\rm BH} = 2.6\times 10^6$ M$_\odot$), while varying the initial conditions for the clusters to explore the future evolution of the galactic nucleus. 
Furthermore, we ran also one simulation in which the galaxy does not host any MBH, in order to highlight 
how a central, massive object can affect the formation of a bright nucleus.

\subsection{The globular cluster system model} 
\label{GCS}

We modeled each of the Henize 2-10 SSCs using a King model \citep{King}, for which 
\cite{ngu14} provides  mass, $M_{\rm SSC}$, core radius, $r_c$, and 
projected galactocentric distance, $R_{\rm SSC}$. We investigate three different possibilities for the initial orbital distribution of the 11 SSCs. 
In the first two models we vary initial conditions for the SSCs, while, in the third case, clusters are distributed as in the first simulation, but the galaxy does not contain a central BH.
The first two initial conditions provide bounds on the dynamical friction inspiral times of the clusters.
In what follows we will refer to these three choices of initial conditions as models S1, S2 and S3.
In model S1 the clusters were assumed to follow the distribution of the background galaxy,
but with the constraint that the clusters projected position (onto a random plane) was consistent 
with the observed values of $R_{\rm SSC}$.
In order to do so we first sample 
the galaxy using the density model of Equation \ref{den}, and then 
selected points from that sample with projected positions (on a random plane) 
similar to those observed for Henize 2-10 SSCs.
The clusters initial velocities were drawn from an isotropic distribution function corresponding to 
the density model of Equation \ref{den}. 
This model represents the one in which the action of the dynamical friction is minimized, since the clusters have initial positions that exceed their projected (observed) positions.
\\
Model S2 corresponds to the assumption that the clusters are all located on the same plane, so that the projected and the spatial distances  coincide, and that they are on circular orbits with velocities as required by the mass distribution of the galaxy. This scenario is motivated by the observed velocities of the clusters that rotate in the same sense as the molecular gas in Henize 2-10, and thus may lie in a disk \citep{santangelo,ngu14}. 
In this case the projected and spatial positions of the clusters coincide, and therefore the effects due to the dynamical friction are the strongest possible.
\\
Finally, for the model S3 we used the same initial conditions as in S1 model, but in this case the galaxy does not contain a central BH. This simulation allows us to study the effect of an SMBH on the NSC formation process.
\\
Once the spatial positions of the clusters were obtained, we evaluated their tidal radius, $r_t$, a parameter needed to define the cluster King model. 
In the hypothesis of circular orbit, the tidal radius of the cluster is given by:
\\
\begin{equation}
r_t^3 = \frac{GM_{\rm SSC}}{\omega^2+(\mathrm{d}^2U/\mathrm{d}r^2)_{r_{\rm SSC}}},
\label{rtid}
\end{equation}
\\
where $\omega$ is the angular velocity of the cluster and $(\mathrm{d}^2U/\mathrm{d}r^2)_{r_{\rm SSC}}$ is the second derivative of the potential evaluated along the SSC circular orbit ($r_{\rm SSC}= const.$).
\\
The knowledge of $r_t$ and $r_c$ makes possible the evaluation of the concentration of the cluster, $c=r_t/r_c$, which correlates with the dimensionless potential well depth, $W_0$, needed to define the King model.
\\
Another parameter we can compare with observations is the effective radius of the cluster, $r_{\rm eff}$, which can be evaluated from $r_c$ through the relation \citep{mioetal}:
\begin{equation}
r_{\rm eff}=(2.43\pm 0.9)r_c.
\end{equation}
The effective radii for the modeled cluster evaluated this way and reported in Table \ref{tab1} are in agreement with those in \cite{ngu14}.
\\
Tables \ref{tab2} and \ref{tab3} provide the main parameters for the star cluster system (SCS) models, S1, S2, and S3 considered in this paper. Note that the cluster parameters for simulations S1 and S3, are the same since the positions of the clusters are the same in these simulations.

\begin{figure}
\centering
\includegraphics[width=8cm]{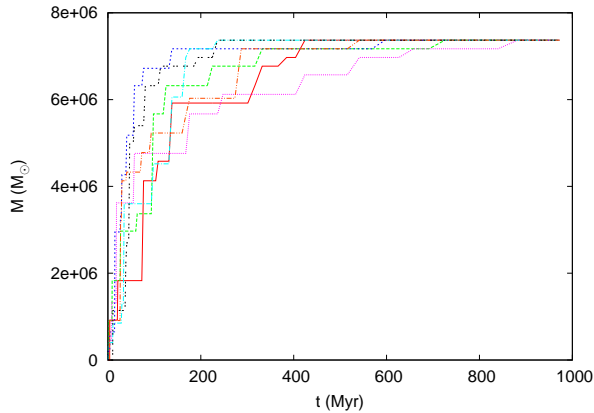}
\caption{Mass accreted around the galactic center as function of time assuming the dynamical friction formula presented in Eq. \ref{tdf}. Different colors correspond to different set of initial conditions for the clusters. The NSC is expected to form within $\sim 500$~Myr.}
\label{growth}
\end{figure}

\begin{table}
\caption{}
\centering{Main parameters of SSCs in configurations S1 and S3.}
\begin{center}
\begin{tabular}{lcrrcrrrr}
\hline
\hline
\multicolumn{1}{l}{ID} & \multicolumn{1}{c}{$W_0$} & \multicolumn{1}{c}{$c$} & \multicolumn{1}{c}{$r_t$} & \multicolumn{1}{c}{$\sigma$} & \multicolumn{1}{c}{$r_{\rm SSC}$}&\multicolumn{1}{c}{$r_{\rm a}$} &\multicolumn{1}{c}{$e $}&\multicolumn{1}{c}{$\tau_{\rm df}$}\\
& & &\multicolumn{1}{c}{(${\rm pc}$)} & \multicolumn{1}{c}{(${\rm km~s^{-1}}$)} & \multicolumn{1}{c}{(${\rm pc}$)}& \multicolumn{1}{c}{(${\rm pc}$)} & &\multicolumn{1}{c}{(${\rm Myr}$)} \\ 
\hline
C1  & $5.3$ & $12.9$ & $16.5$ & $24.5$ & $87.9$&$145$&$0.4$&$43.4$\\  
C2  & $5.5$ & $14.5$ & $11.2$ & $18.8$ & $71.6$&$75$&$0.3$&$28.4$\\  
C3  & $5.1$ & $11.7$ & $11.7$ & $20.5$ & $67.6$&$200$&$0.8$&$85.3$\\
C4  & $4.1$ & $7.4$  & $5.3$  & $27.3$ & $12.4$&$30$&$0.4$&$7.1$\\
C5  & $6.1$ & $19.1$ & $10.1$ & $13.1$ & $109.1$&$120$&$0.5$&$116.2$\\
C6  & $6.2$ & $22.3$ & $11.8$ & $12.1$ & $148.1$&$190$&$0.5$&$270.0$\\
C7  & $5.2$ & $12.3$ & $10.9$ & $13.5$ & $116.4$&$140$&$0.5$&$138.8$\\
C8  & $6.5$ & $25.2$ & $11.6$ & $12.9$ & $132.5$&$160$&$0.6$&$163.4$\\
C9  & $6.2$ & $21.0$ & $7.8$  & $10.5$ & $101.6$&$180$&$0.5$&$377.4$\\
C10 & $6.1$ & $19.2$ & $8.8$  & $14.8$ & $72.2$&$180$&$0.6$&$203.0$\\
C11 & $5.9$ & $17.7$ & $6.5$  & $11.5$ & $66.6$&$140$&$0.4$&$265.4$\\
\hline
\end{tabular}
\end{center}
\begin{tablenotes}
	\item Column 1: cluster name. Column 2: adimensional central potential. Column 3: concentration parameter. Column 4: tidal radius. Column 5: velocity dispersion. Column 6: galactocentric distance. Column 7: initial apocentric distance of the SSC orbit. Column 8: eccentricity of the orbit. Column 9: dynamical friction time evaluated using Eq. \ref{tdf}.
\end{tablenotes}
\label{tab2}
\end{table}

\begin{table}
\caption{}
\centering{Main parameters of SSCs in configuration S2.}
\begin{center}
\begin{tabular}{lcrrcrrrr}
\hline
\hline
\multicolumn{1}{l}{ID} & \multicolumn{1}{c}{$W_0$} & \multicolumn{1}{c}{$c$} & \multicolumn{1}{c}{$r_t$} & \multicolumn{1}{c}{$\sigma$} & \multicolumn{1}{c}{$r_{\rm SSC}$}&\multicolumn{1}{c}{$r_{\rm a}$} &\multicolumn{1}{c}{$e $}&\multicolumn{1}{c}{$\tau_{\rm df}$}\\
& & &\multicolumn{1}{c}{(${\rm pc}$)} & \multicolumn{1}{c}{(${\rm km~s^{-1}}$)} & \multicolumn{1}{c}{(${\rm pc}$)}& \multicolumn{1}{c}{(${\rm pc}$)} & &\multicolumn{1}{c}{(${\rm Myr}$)} \\ 
\hline
C1  & $5.1$ & $11.5$ & $14.6$ & $26.1$ & $ 64.3$ &$64.3$&$0$&$20.0$\\
C2  & $5.1$ & $11.6$ & $ 8.9$ & $21.1$ & $ 37.0$ &$37.0$&$0$&$13.9$\\
C3  & $4.2$ & $ 7.6$ & $ 7.6$ & $25.3$ & $ 20.5$ &$20.5$&$0$&$4.3$\\
C4  & $4.0$ & $ 6.9$ & $ 4.9$ & $28.2$ & $ 11.3$ &$11.3$&$0$&$1.7$\\
C5  & $5.9$ & $17.1$ & $ 9.1$ & $13.8$ & $ 85.6$ &$85.7$&$0$&$106.7$\\
C6  & $5.8$ & $16.5$ & $ 8.7$ & $14.0$ & $ 78.0$ &$78.0$&$0$&$90.6$\\
C7  & $5.2$ & $12.2$ & $10.8$ & $13.5$ & $114.2$ &$114.2$&$0$&$161.3$\\
C8  & $6.4$ & $25.0$ & $11.5$ & $13.0$ & $131.3$ &$131.3$&$0$&$209.3$\\
C9  & $6.0$ & $18.4$ & $ 6.8$ & $11.2$ & $ 74.5$ &$74.5$&$0$&$132.9$\\
C10 & $6.1$ & $19.0$ & $ 8.8$ & $14.9$ & $ 70.9$ &$70.9$&$0$&$70.7$\\
C11 & $5.8$ & $16.6$ & $ 6.1$ & $11.8$ & $ 56.1$ &$56.1$&$0$&$80.7$\\
\hline
\end{tabular}
\end{center}
\begin{tablenotes}
	\item Column 1: cluster name. Column 2: adimensional central potential. Column 3: concentration parameter. Column 4: tidal radius. Column 5: velocity dispersion. Column 6: galactocentric distance. Column 7: initial apocentric distance of the SSC orbit. Column 8: eccentricity of the orbit. Column 9: dynamical friction time evaluated using Eq. \ref{tdf}.
\end{tablenotes}
\label{tab3}
\end{table}

\subsection{The NSC formation process in Henize 2-10: analytical predictions}
\label{semia}

Dynamical friction leads star clusters to decay toward the center of their parent galaxies (see for example \cite{TrOsSp}, \cite{Dolc93}). In general, the easiest way to describe the action of dynamical friction in a galactic environment is by means of  Chandrasekhar's formula \citep{Cha43I} in its local approximation. However, many works \citep{OBS,Pes92,RCD05,ASCD14} have shown
that this local approximation fails when the satellite moves in the innermost region of a galaxy.
\\
\cite{ASCD14} developed a useful formula for the dynamical friction decay time valid for both cored and cuspy density profiles, and for galaxies with a central black hole:
\\
\begin{equation}
\tau_{\rm df} (\mathrm{Myr}) = 0.3 \sqrt{\frac{r_s^3}{M_{\rm H}}}g(e,\gamma)\left(\frac{M_{\rm H}}{M_{\rm SSC}}\right)^{0.67}\left(\frac{r_{\rm a}}{r_s}\right)^{1.76},
\label{tdf}
\end{equation}
\\
where $\gamma$ is the previously defined exponent in the galactic density profile, $0\leq e\leq 1$ is the orbital eccentricity of the point-like object whose mass is $M_{\rm SSC}$, $r_{\rm a}$ is the orbit apocenter. 
Here, we used for the eccentricity the definition:
\begin{equation}
e = \frac{r_{\rm a}-r_{\rm p}}{r_{\rm a}},
\end{equation}
being $r_{\rm a}$ and $r_{\rm p}$ the apocentral and pericentral distance, respectively.
Parameters $r_s$ (in kpc) and $M_{\rm H}$ (in $10^{11}\rm M_\sun$) are the scale radius and the total mass of the galaxy $\gamma$ model. 
For the scopes of this paper, we performed several new direct N-body runs to test the validity of the formula for massive objects on circular and quasi radial orbits and for $\gamma$ values smaller than $2$, finding that a function $g(e,\gamma)$ that gives a very good fit to results has this expression:
\begin{equation}
g(e,\gamma)=(2-\gamma)\left[a_1\left(\frac{1}{(2-\gamma)^{a_2}}+a_3\right)(1-e) + e\right],
\end{equation}
with $a_1= 2.63 \pm 0.17$, $a_2 = 2.26 \pm 0.08$ and $a_3=0.9 \pm 0.1$. The expression for $g(e,\gamma)$ here is slightly different from the one found in \cite{ASCD14} due to that that paper the authors mainly focused their attention on the study of the dynamical friction process in models with $\gamma\leq 1$ for both circular and radial orbits, and $\gamma > 1$ limiting to nearly radial orbits. Here, we have extended the interval of validity of the fitting formula to circular and radial simulations in the range $0\leq \gamma<2$.
Note, however, that Equation \ref{tdf} is valid only for $\gamma<2$.
\\
Equation \ref{tdf} can be used to estimate the mass accumulation around the center of Henize 2-10 as a consequence of the orbital decay of the 11 clusters. 
\\
To have a meaningful idea of what we should expect from $N$-body simulations, for each of the 11 SSCs in Henize 2-10 we randomly {\textbf selected} initial orbital eccentricity and position values. Then, for any given set of initial parameters we evaluated the mass accreted to the galaxy center.
\\
Figure \ref{growth} shows the mass accumulated around the center of the galaxy for 7 sets of random initial conditions: a mass of $\sim 7\times$10$^6$~M$_\odot$ is accreted into the center of the galaxy with timescales between $0.1$ and $0.5$ Gyr. This mass is consistent with the mass of observed NSCs in other galaxies \cite[e.g.][]{seth08}.
\\
Tables \ref{tab2} and \ref{tab3} give the decay time of each cluster in the configuration S1 and S2, respectively, evaluated using Equation \ref{tdf}. 
\\
The semi-analytical formula of Equation \ref{tdf} neglects the mass loss of clusters along their orbits, and thus represents a lower limit to the time needed to the decay and an upper limit to the quantity of cluster mass deposited into a NSC. A more realistic treatment requires N-body simulations, which we present in the following sections.

\begin{table}
\caption{}
\centering{Estimated initial galactocentric distances of the SSCs for different assumptions of their ages.}
\begin{center}
\begin{tabular}{crrr}
\hline
\hline
\multicolumn{1}{c}{ID}  & &\multicolumn{1}{c}{$r_0/r_{\rm SSC}$ }& \\
\hline
C$1$&$1.1$&$1.3$&$2.8$\\
C$2$&$1.2$&$1.4$&$3.3$\\
C$3$&$1.6$&$2.0$&$6.1$\\
C$4$&$2.2$&$3.0$&$10.1$\\
C$5$&$1.0$&$1.1$&$1.5$\\
C$6$&$1.0$&$1.1$&$1.5$\\
C$7$&$1.0$&$1.0$&$1.3$\\
C$8$&$1.0$&$1.0$&$1.2$\\
C$9$&$1.0$&$1.0$&$1.4$\\
C$10$&$1.0$&$1.1$&$1.7$\\
C$11$&$1.0$&$1.1$&$1.6$\\
\hline
\end{tabular}
\end{center}
\begin{tablenotes}
\item Column 1: cluster name. Column 2-4: ratio between the initial galactocentric distance the cluster had at its birth, $r_0$, and its current position, $r_{SSC}$, assuming as age estimates: $5$, $10$ and $100$ Myr, respectively.
\end{tablenotes}
\label{ag}
\end{table}

The work by \cite{chandar03} suggests that all the 11 SSCs studied are very young, $\tau \simeq 5$ Myr, therefore their initial galactocentric distances, $r_0$, are likely very close to the presently observed, $r_{SSC}$, values.
We can quantify this by mean of an easy inversion of Equation \ref{tdf}, which tells us how much the SSCs orbits have shrunk over their (short) life. The data Table \ref{ag} are obtained in the simplifying assumption of SSCs on initially circular orbits with following circular orbital  shrink. 
Data in Table \ref{ag} indicate, even in the very unlikely case that the clusters age reaches $100$ Myr, they should have formed within $1$ kpc from the center of Henize 2-10.

\section{Results from \textit{N}-body simulations}
\label{nbody}
\subsection{The ``number of particles'' problem}
\label{numbprob}

In this work we simulate the future dynamical evolution of the clusters observed in Henize 2-10, to see whether or not a nuclear stellar cluster will form, and what the timescale of this formation process is.
To do this, we used the direct $N$-body code HiGPUs \citep{Spera}, a $6^{th}$-order Hermite integrator that runs on hybrid platforms where CPUs are coupled to Graphic Processing Units (GPUs) acting as computing accelerators, fully exploiting parallelization.

Despite the great accuracy and speed of the code, the number of particles that we can use in this kind of simulations is limited to  $\sim 10^6$. 
Since Henize 2-10 has a mass of 10$^{10}$~M$_\odot$, it is clearly impossible to create a reliable, star-by-star simulation of the whole galaxy.
To address this problem, we limit the extension of our $N$-body sampling of the galaxy to a relatively small region, as explained in Section \ref{galmod}. 
In the simulations we set a gravitational softening $\epsilon = 0.02$ pc, which is a value much smaller than the core radius of the clusters, and thus allows for a good description of the internal dynamics of the clusters.

Moreover, we allowed for a difference between the mass of particles representing the galaxy, $m_{g}$, and that of the particles in the clusters, $m_{\rm SSC}$, such that $m_{g}/m_{\rm SSC}=8$. 
This allows us to have a number of particles in the clusters,
with even the smallest cluster having $N\simeq 2000$, sufficient to ensure reliable results on their dynamical evolution. Specifically, the two-body relaxation time of the smallest cluster is long enough to make the evaporation time $T_{\rm evp} \simeq 140 T_{\rm rel}\simeq 2.5$ Gyr \citep{bt}, much longer than the run time of our $N$-body simulations. This guarantees that the global properties of the small clusters are preserved along the integration time despite being undersampled.

\subsection{Configuration S1}
\begin{figure*}
\subfigure{
\includegraphics[width=8cm]{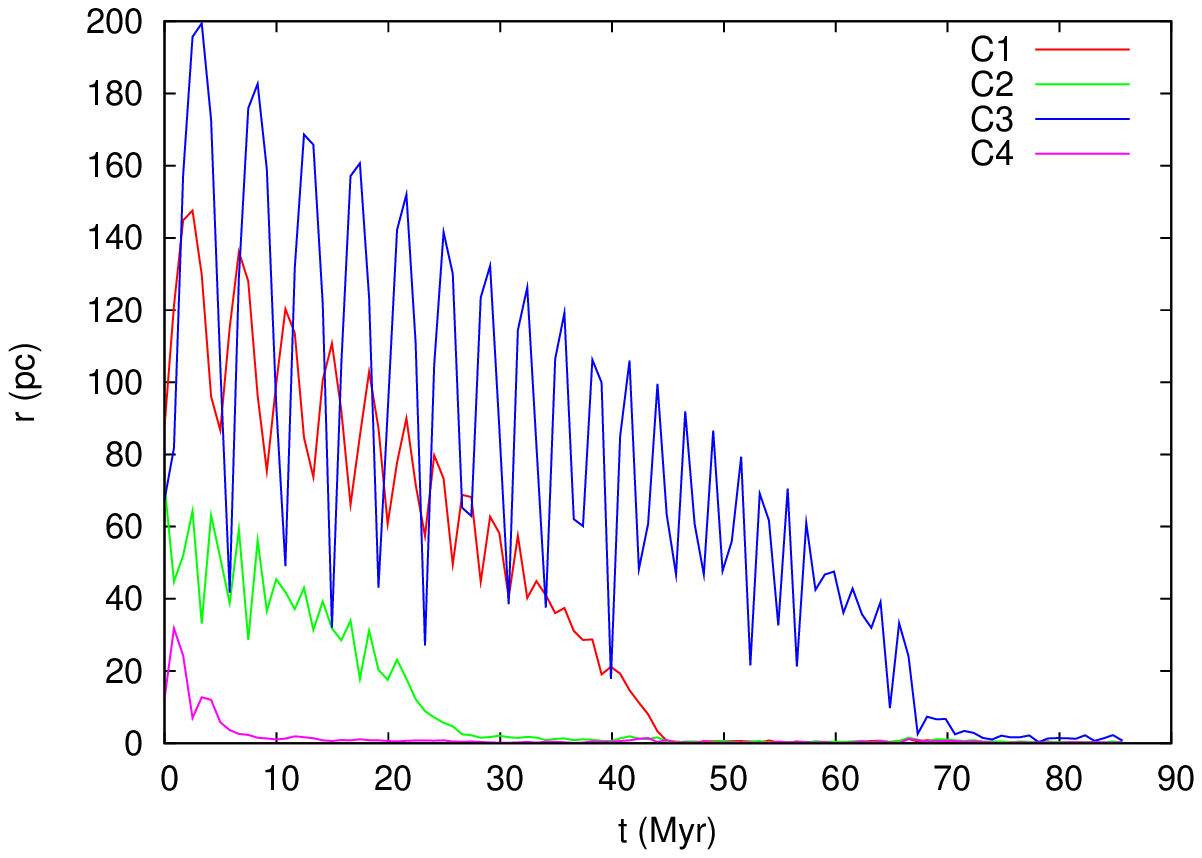}
\includegraphics[width=8cm]{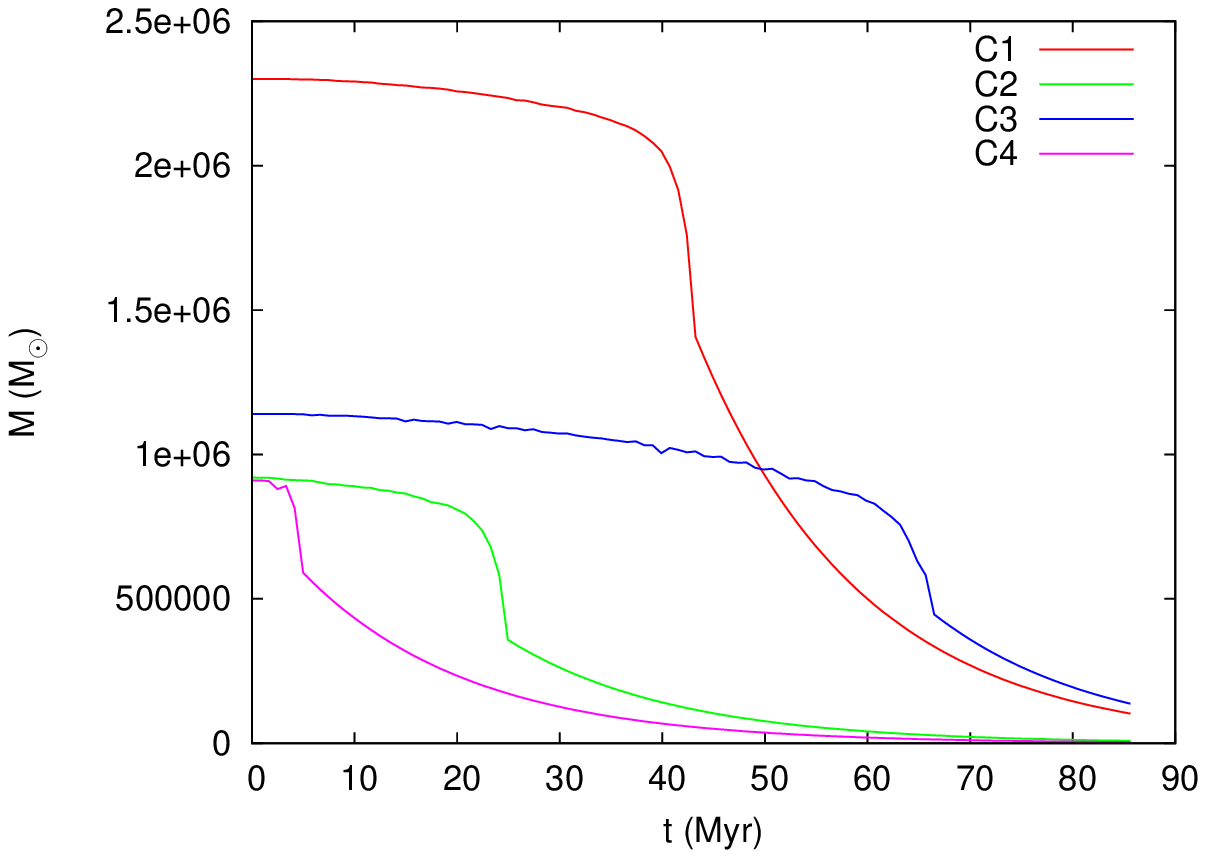}
}
\subfigure{
\includegraphics[width=8cm]{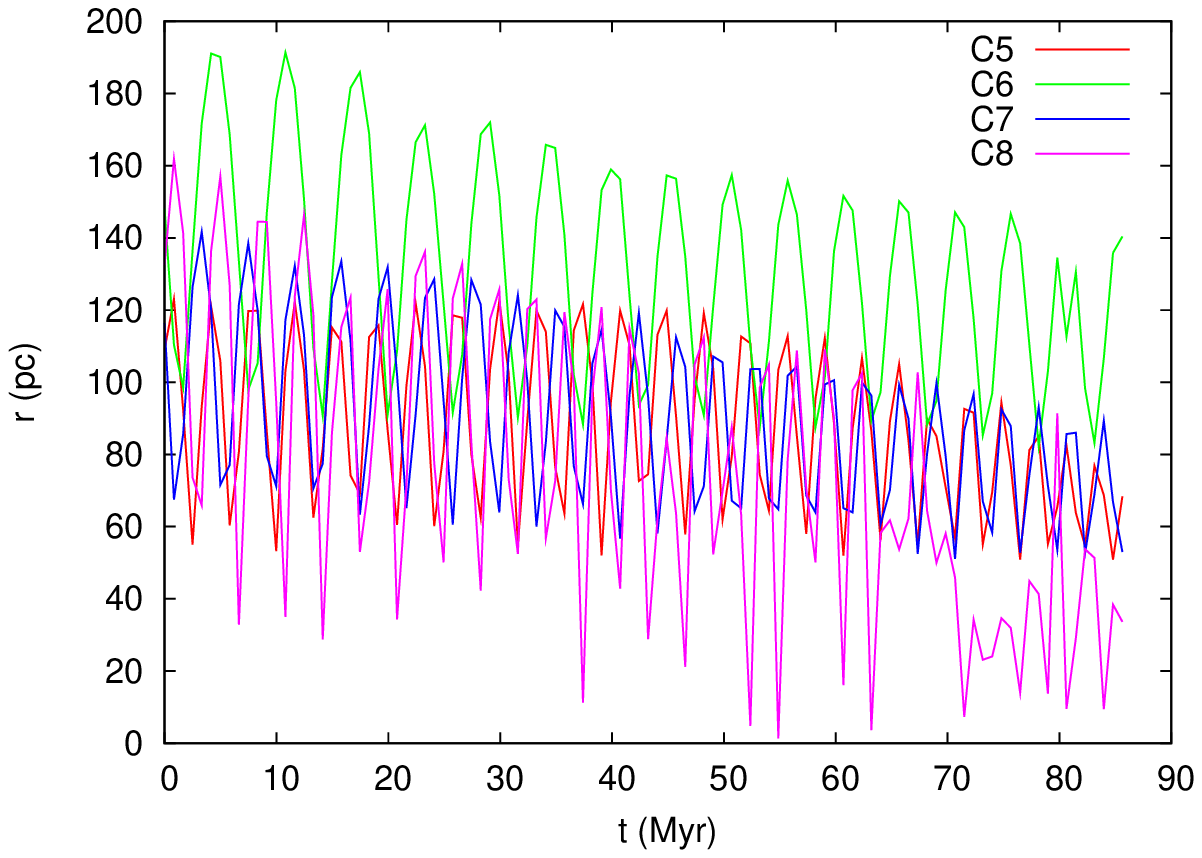}
\includegraphics[width=8cm]{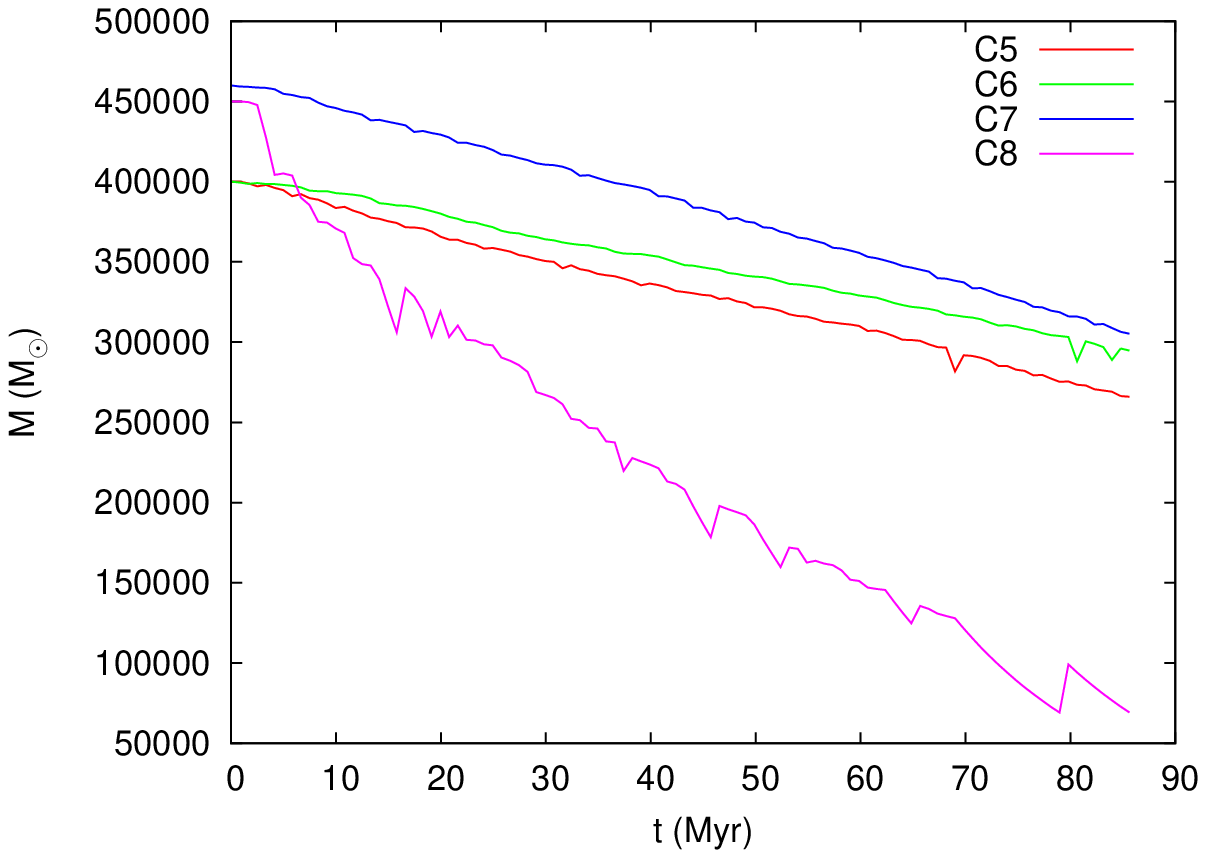}
}
\subfigure{
\includegraphics[width=8cm]{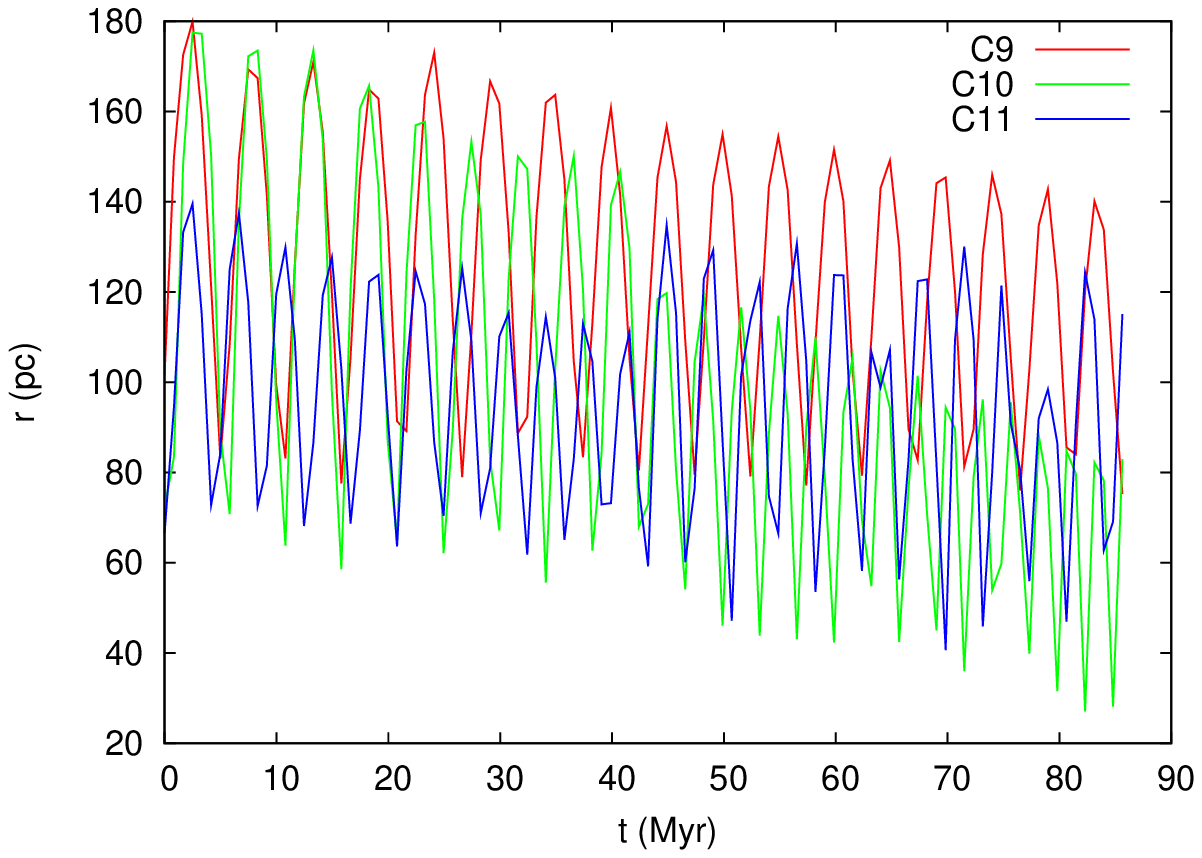}
\includegraphics[width=8cm]{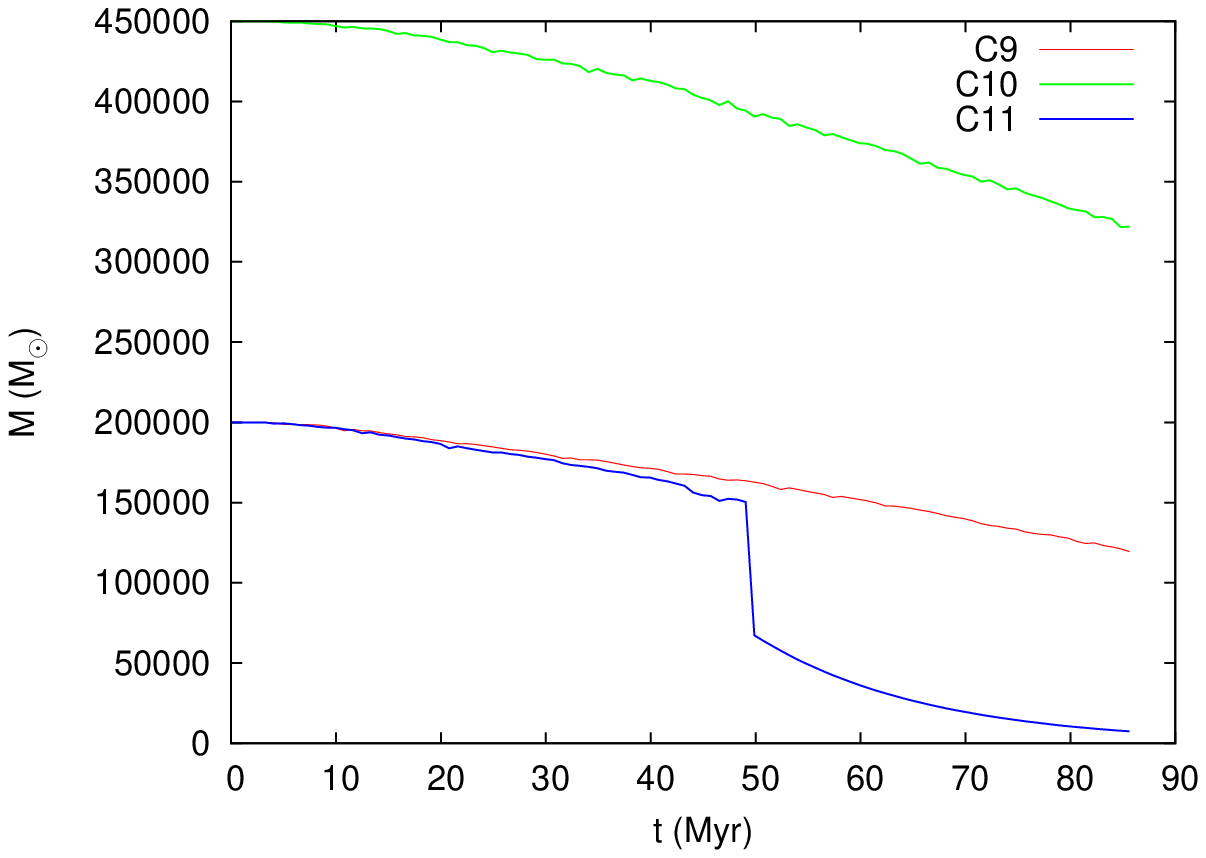}
}
\caption{Left panels: radial distance of each cluster from the BH as a function of time. Right panels: mass of each cluster as function of time. It is evident that the main contributors to the nuclear mass are clusters C1-C4, which transport a significant amount of their initial mass to the center. Panels refer to simulation S1}
\label{traj1}
\end{figure*} 

As explained above, the clusters in the S1 configuration have initial conditions obtained directly from the distribution function of the galaxy model given in Equation \ref{den}.

Table~\ref{tab2} summarizes the parameters that describe the SCS in this case.

In the following, each cluster will be identified with the letter C and a number between 1 and 11, as in Table~\ref{tab1}.

Figure~\ref{traj1} shows the distance of each cluster from the central galactic BH as a function of time, as obtained by our $N$-body simulation. 
It is worth noting that an error on the position of the galaxy center would not affect our estimates, since we are interested in the mutual distances between the clusters and the BH.
As expected, while the most massive clusters (C1-C4) reach the BH within $\sim 50$ Myr, the dynamical friction is less efficient on lighter clusters, which tend to remain on their initial orbits for longer times. 

\begin{figure}
\centering
\includegraphics[width=8cm]{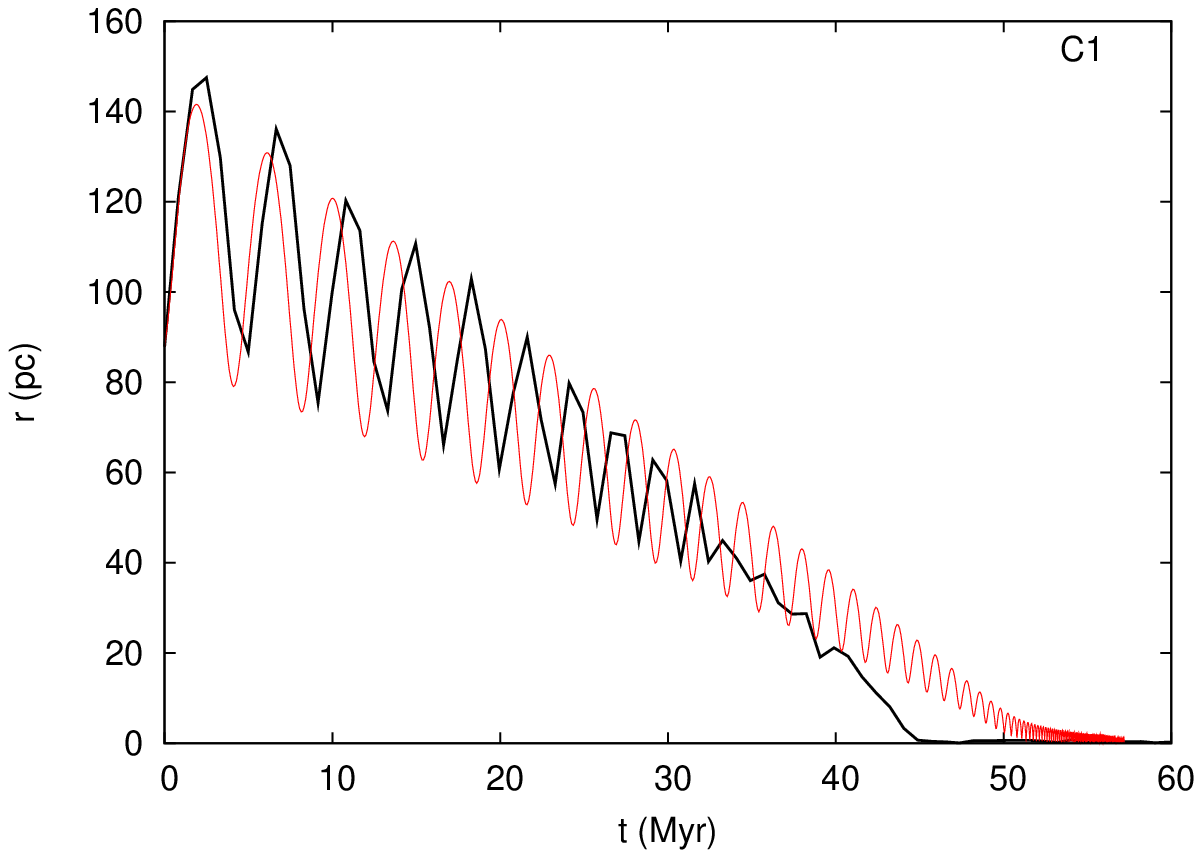}\\
\includegraphics[width=8cm]{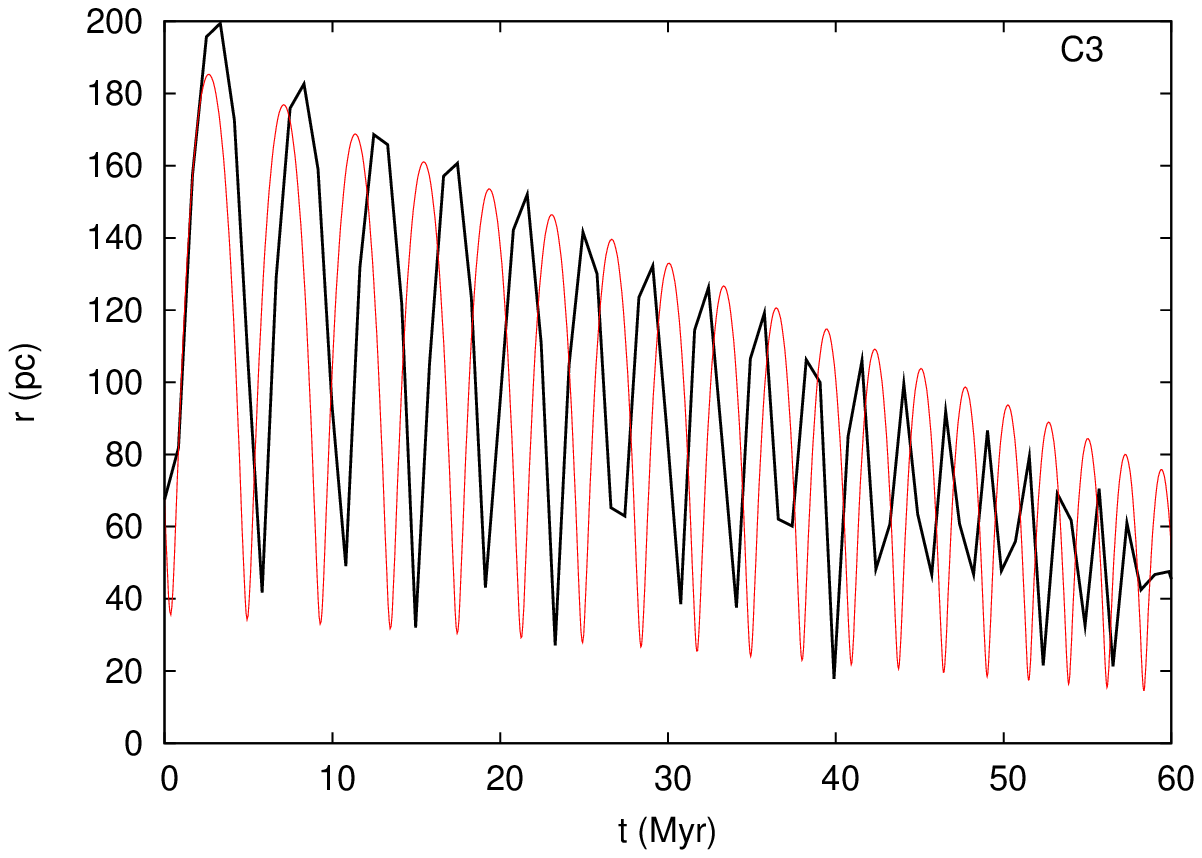}\\
\includegraphics[width=8cm]{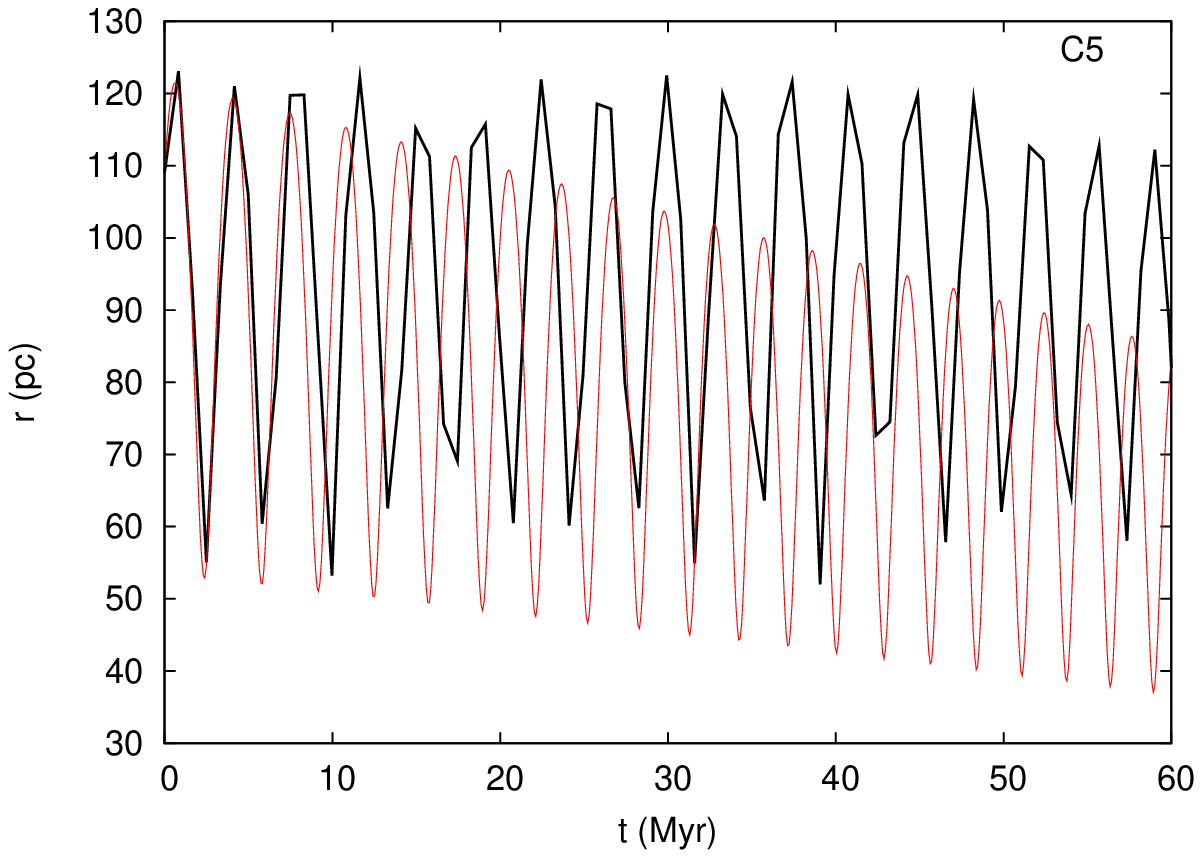}\\
\caption{From top to bottom: radial distance from the BH of the cluster C1, C3 and C5 as a function of time obtained with the self-consistent simulation S1 (thick black line), and the semi-analytical method described in \cite{ASCD14} (thin red line). The comparison shows a good agreement for heavy clusters, while the analytical description is in poorer agreement for smaller clusters.}
\label{cmptraj}
\end{figure}

Figure~\ref{cmptraj} shows that the time evolution of the cluster orbits obtained through the $N$-body simulation agrees with the semi-analytical method described in \cite{ASCD14}. 
It is evident that the semi-analytical orbit accurately describe the cluster orbital evolution for the most massive clusters, confirming the reliability of Equation~\ref{tdf}. On the other hand, the motion of lighter clusters is strongly affected by the tidal forces induced by the BH and the other clusters. Therefore, in this case the semi-analytical approach is less accurate, but even for these clusters the results are in broad agreement with the $N$-body results.

As the clusters move within the galaxy, their initial spherical shape changes significantly due to the action of the tidal forces. In this case, it is very difficult to define general quantities of the clusters such as their center or mass. To give an estimate of the cluster mass at any time, we obtained $r_{t}$ solving Equation \ref{rtid}, while the position of the cluster center is estimated through a grid-based algorithm. Once the tidal radius and cluster center have been evaluated, we considered the cluster mass in a given time-step to be the mass enclosed within $r_{t}$.

In the right panels of Figure \ref{traj1} the evolution in time of the mass of the clusters is shown.  In this case, it is evident that massive clusters (C1-C4), lose their mass when they reach the galaxy center, contributing significantly to the formation of the nucleus.

On the other hand, small clusters (C5-C10) lose mass continuously throughout the simulation to the field.
The evolution of cluster C8 and C11 are significantly affected by tidal forces, which strip away more than $70 \%$ of the initial clusters mass well before they reach the center of the galaxy.

Combining the informations on the time evolution of the SSCs mass and trajectory, we can extrapolate the amount of mass that each cluster will 
carry to the innermost region of the galaxy. In particular, Table \ref{tab7} shows the percentage of mass dragged within $20$ pc from the galactic center for all the simulations performed. 
In the case of configuration S1, it is evident that tidal forces should prevent the complete decay of small clusters, disrupting completely clusters C5-C9 and C11, and leaving cluster C10 with only $\sim 30\%$ of its initial mass.

\begin{table}
\caption{}
\centering{Percentage of mass left to the galactic center by each cluster in the three configurations studied.}
\begin{center}
\begin{tabular}{crrr}
\hline
\multicolumn{1}{c}{ID} & \multicolumn{1}{c}{S1} & \multicolumn{1}{c}{S2} & \multicolumn{1}{c}{S3}  \\
\hline
\hline
C1  &  $ 80$ & $80$ & $93$ \\
C2  &  $ 80$ & $95$ & $89$ \\
C3  &  $ 70$ & $95$ & $73$ \\
C4  &  $ 95$ & $95$ & $98$ \\
C5  &  $ 0$  & $30$ & $ 0$ \\
C6  &  $ 0$  & $40$ & $ 0$ \\
C7  &  $ 0$  & $40$ & $ 0$ \\
C8  &  $ 0$  & $10$ & $ 0$ \\
C9  &  $ 0$  & $ 0$ & $ 0$ \\
C10 &  $ 30$ & $70$ & $30$ \\
C11 &  $ 0$  & $60$ & $ 0$ \\
\hline
\end{tabular}
\end{center}
\begin{tablenotes}
\item Column 1: cluster name. Column 2-4: percentage of the cluster mass left within $20$ pc from the BH in configuration S1, S2 and S3, respectively.
\end{tablenotes}
\label{tab7}
\end{table}

\begin{figure}
\centering
\includegraphics[width=8cm]{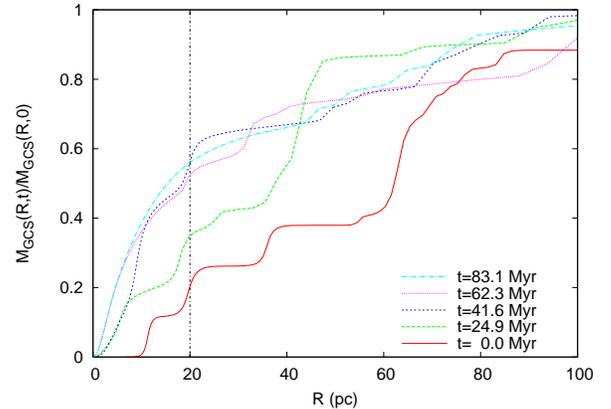}
\caption{The mass distribution, as function of the projected radial distance from the BH, of the SCS at different times, normalized to the total SCS mass for configuration S1. The mass enclosed within $20$ pc reaches a nearly constant value after $40$ Myr.}
\label{mvsr}
\end{figure}

Figure \ref{mvsr} gives information on the mass transported to the galactic center, showing the cumulative radial mass distribution of the SCS at different times. 
It is evident an initial phase lasting $\sim 40$ Myr in which the mass accumulated within $20$ pc increases until it reaches $\sim 55\%$ of the total SCS mass.  This represents in this configuration a ``saturation'' value for the accumulated mass. Combining this figure with Figure \ref{traj1} reveals that the major contribution to the central mass is from the C1-C4 clusters, while the lighter clusters decay more slowly. 
Moreover, tidal forces act more efficiently on small clusters, leading to mass loss; this makes the decay of these clusters even slower.

\begin{figure}
\centering
\includegraphics[width=8cm]{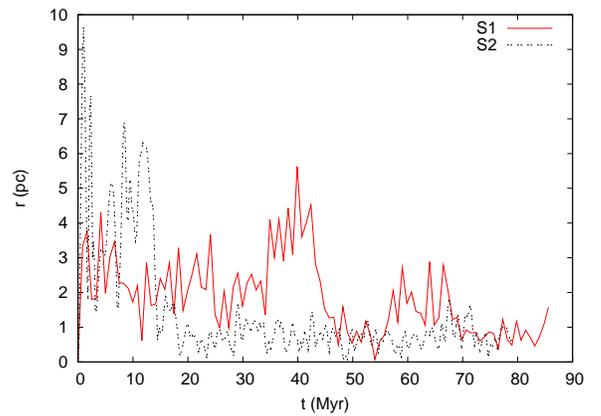}
\caption{Evolution of the distance of the BH from the galaxy center as function of time in cases S1 and S2.}
\label{BHt}
\end{figure}

\begin{figure}
\centering
\includegraphics[width=8cm]{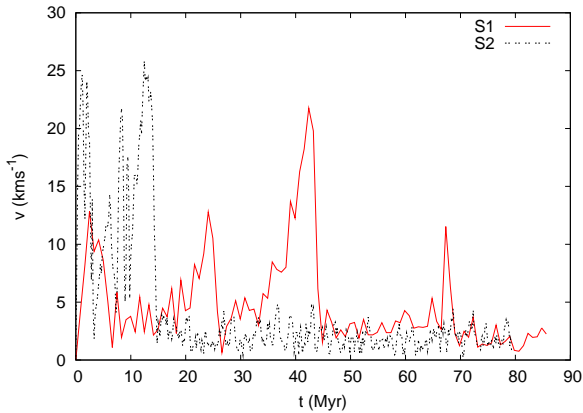}
\caption{Time evolution of the BH speed in cases S1 and S2.}
\label{BHvel}
\end{figure}

Figure~\ref{BHt} shows the variation with time of the distance of the BH to the galactic center, revealing that in this S1 case the BH motion remains limited to a radial region of  $<$6 pc from its initial position. The maximum displacement, at about $40$ Myr, corresponds to the close interaction with C1 (the heaviest cluster), as shown in a comparison of Figure \ref{BHt} and Figure \ref{traj1}.
After the completion of the main mergers, the BH oscillates within 1 pc from the center with a residual speed of $\sim 2$ km s$^{-1}$ (see Figure \ref{BHvel}).

\begin{figure}
\centering
\includegraphics[width=8cm]{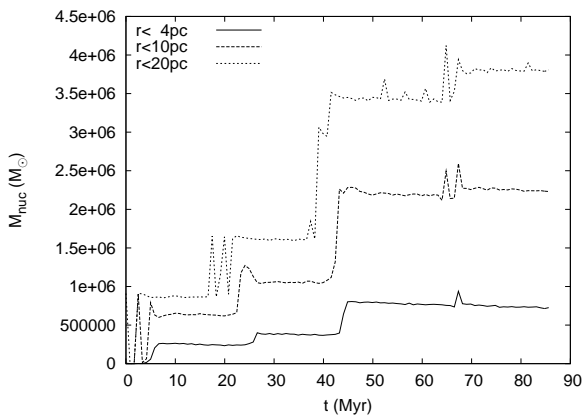}
\caption{Mass deposited in a sphere centered in the BH of radius: $4$ pc (straight line), $10$ pc (dashed line), and $20$ pc (dotted line), for the simulation S1.}
\label{NSCg}
\end{figure}

To understand whether a NSC forms, we evaluated the cluster mass deposited in the innermost region of the galaxy.
This is shown in Figure \ref{NSCg}, which shows the amount of cluster mass within $4$, $10$ and $20$ pc distance from the central BH as a function of time.

The evident steps in the increasing form of the function flag the time at which a new cluster reaches the Henize center. After $90$ Myr, the deposited mass within $20$ pc reaches a nearly constant value of $4\times 10^6$ M$_\odot$, i.e. $51\%$ of the total SCS mass.

The presence of a NSC in the center of a galaxy is, observationally, usually revealed by the study of the surface luminosity profile of the inner part of the galaxy. In fact, a NSC tipically emerges in such profiles as an overdensity in the innermost region. 

\begin{figure}
\centering
\includegraphics[width=8cm]{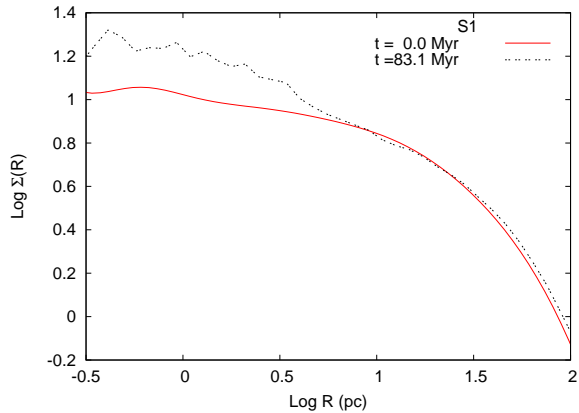}
\caption{Surface density profile of the system galaxy+SCS at time $t=0$ Myr (straight line) and $t=83.1$ Myr (dotted line) for the configuration S1. The x-axis is the projected distance from the galaxy center.} 
\label{supd1}
\end{figure}

The initial and final surface density profiles are shown in Figure \ref{supd1}. It shows clearly the formation of a well-visible structure which extends up to $0.2"$ ($10$ pc). Since the detection of a NSC is made observationally by evaluating the mass contained within the overdensity, our newly born NSC has an observational extension of $10$ pc, a mass of $M_{\rm NSC}=4.6\times 10^6$ M$_\odot$ and an effective radius $r_{\rm NSC}\simeq 4.17$ pc. 
It should be highlighted that, here, $M_{NSC}$ is obtained summing the mass which the decayed clusters carried to the galactic center and the mass of the galactic background enclosed within $10$ pc from the BH.
\\
On the other hand, the mass deposited through the cluster decay in the range $10-20$ pc ($\lesssim 2\times 10^6$ M$_\odot$) is evidently not dense enough to give rise to a clearly detectable structure.

\subsection{Configuration S2}

In the configuration S2 the SSC orbits lie on the plane perpendicular to the line of sight. 
Such a configuration would result in the strongest dynamical friction.  This scenario may also be closest to the clusters true distribution.  Indeed, the young clusters, all located to the East of the black hole and dynamical center, are all blueshifted  \citep{ngu14}.  The gas in this region is rotating in the same direction \citep{cresci}, and overall the gas in the galaxy has a low inclination, $i \sim$38$^\circ$ \citep{kobul95}. Thus it appears reasonable that the young clusters are in a rotating disk close to the plane of the sky.

Parameters of the clusters in this configuration are summarized in Table \ref{tab3}.
 
In this case, clusters reach the galactic center in shorter times than in case S1 since they move on projected orbits, which are of course smaller than the corresponding 3D positions. 

The exceptions are clusters C7 and C8, that move as a binary. 

\begin{figure*}
\subfigure{
\includegraphics[width=8cm]{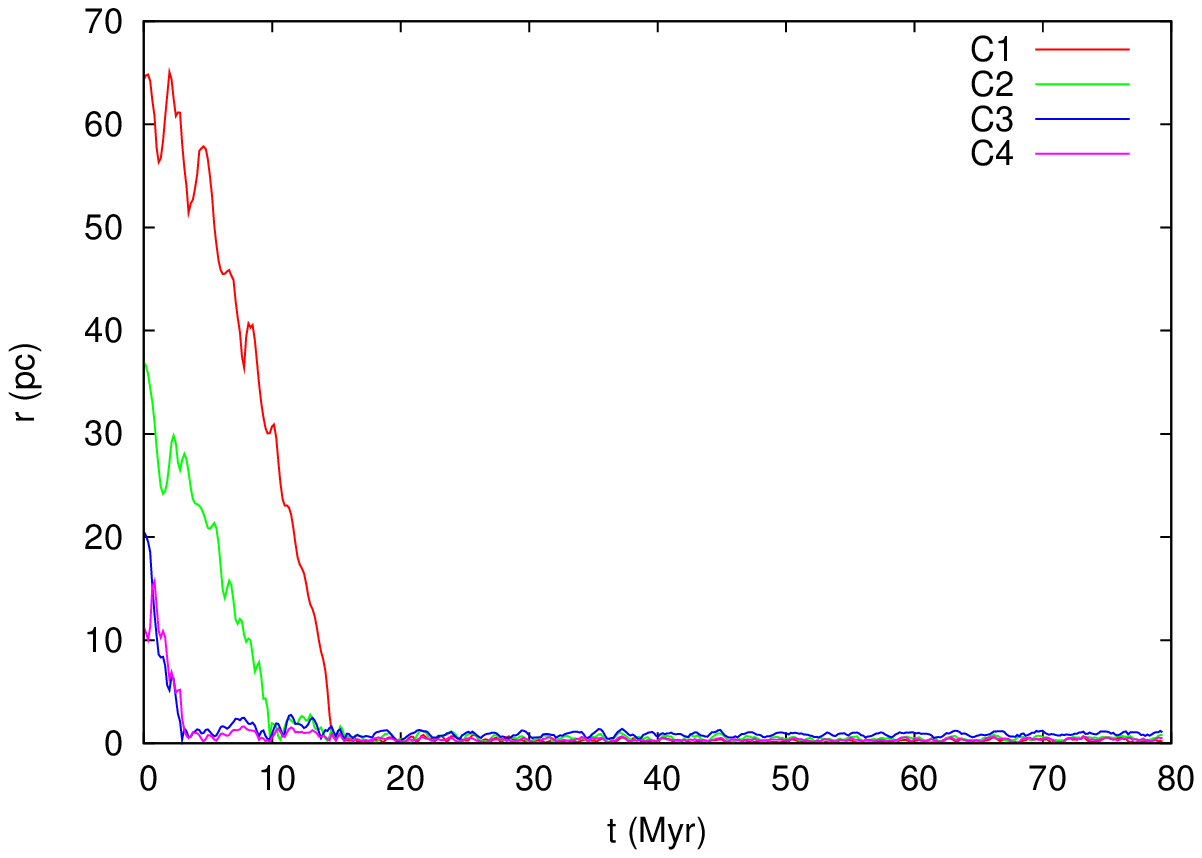}
\includegraphics[width=8cm]{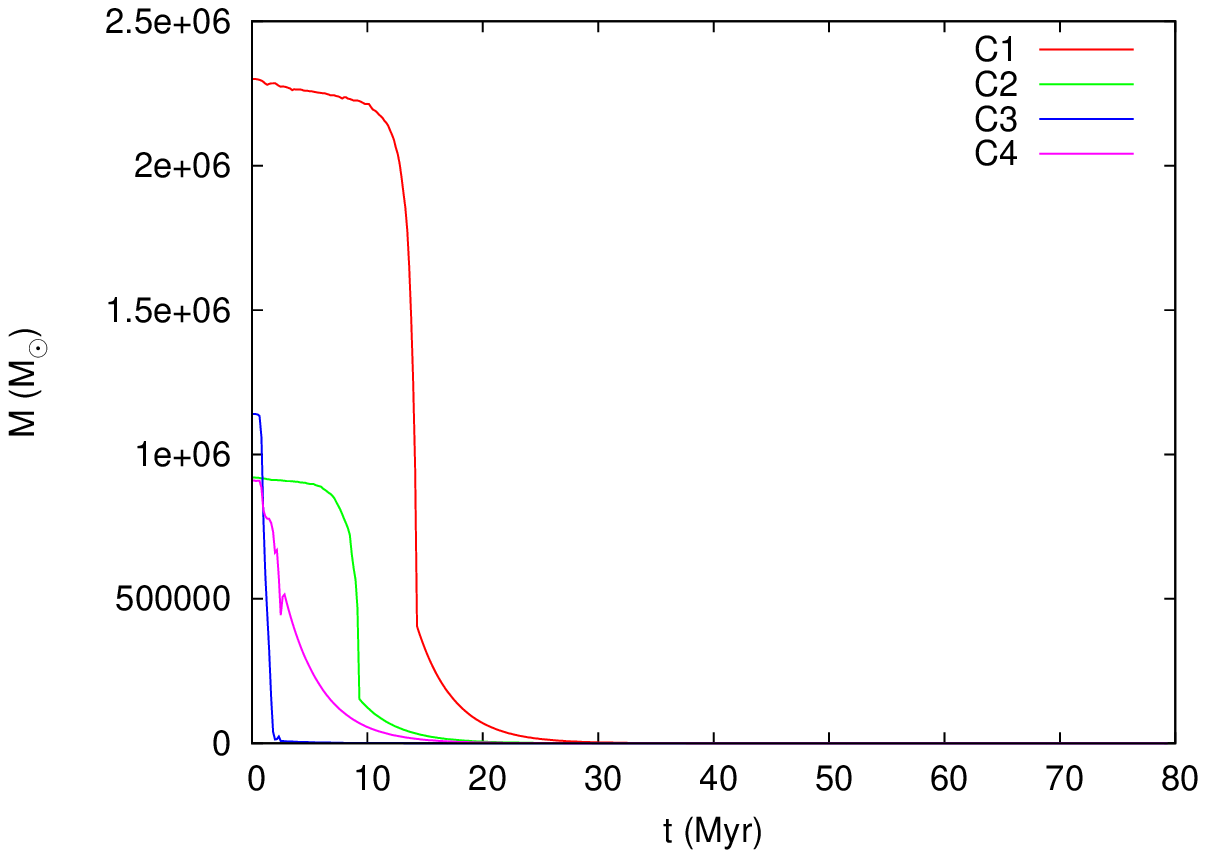}
}
\subfigure{
\includegraphics[width=8cm]{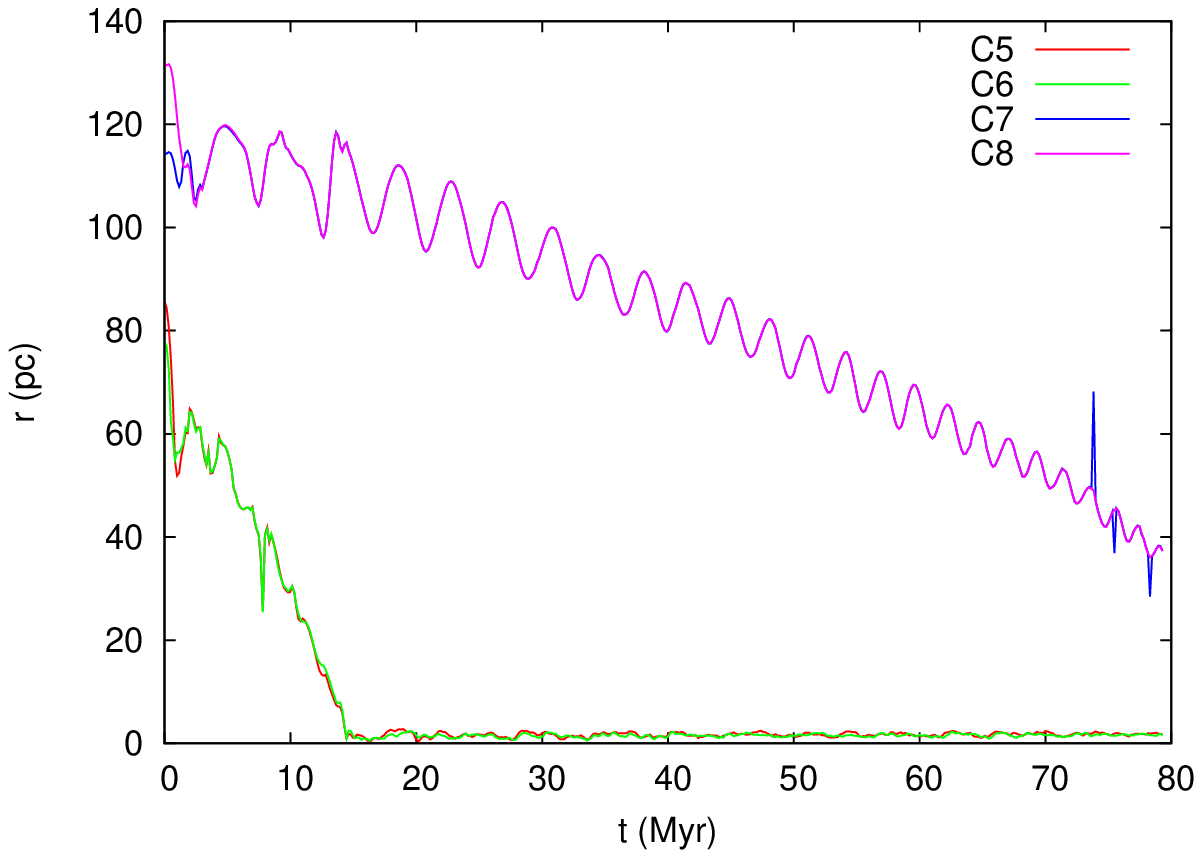}
\includegraphics[width=8cm]{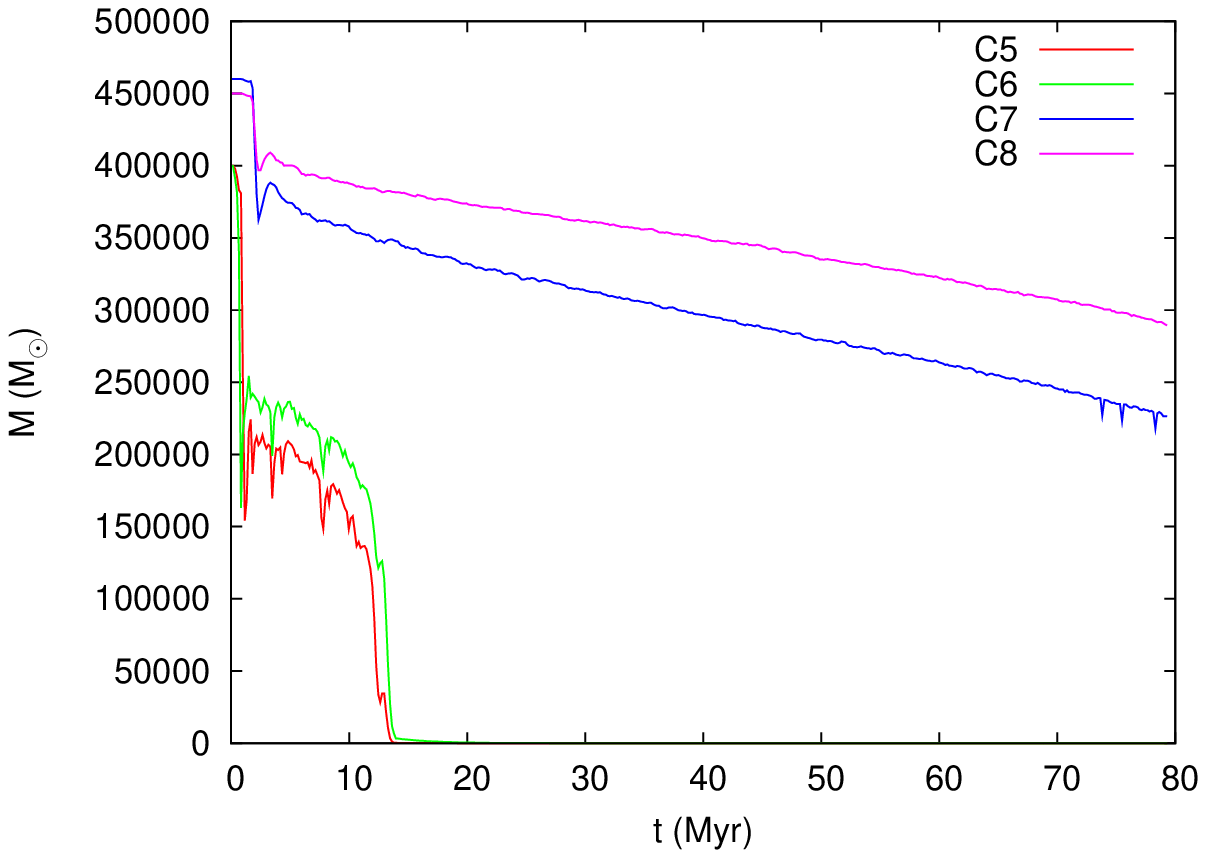}
}
\subfigure{
\includegraphics[width=8cm]{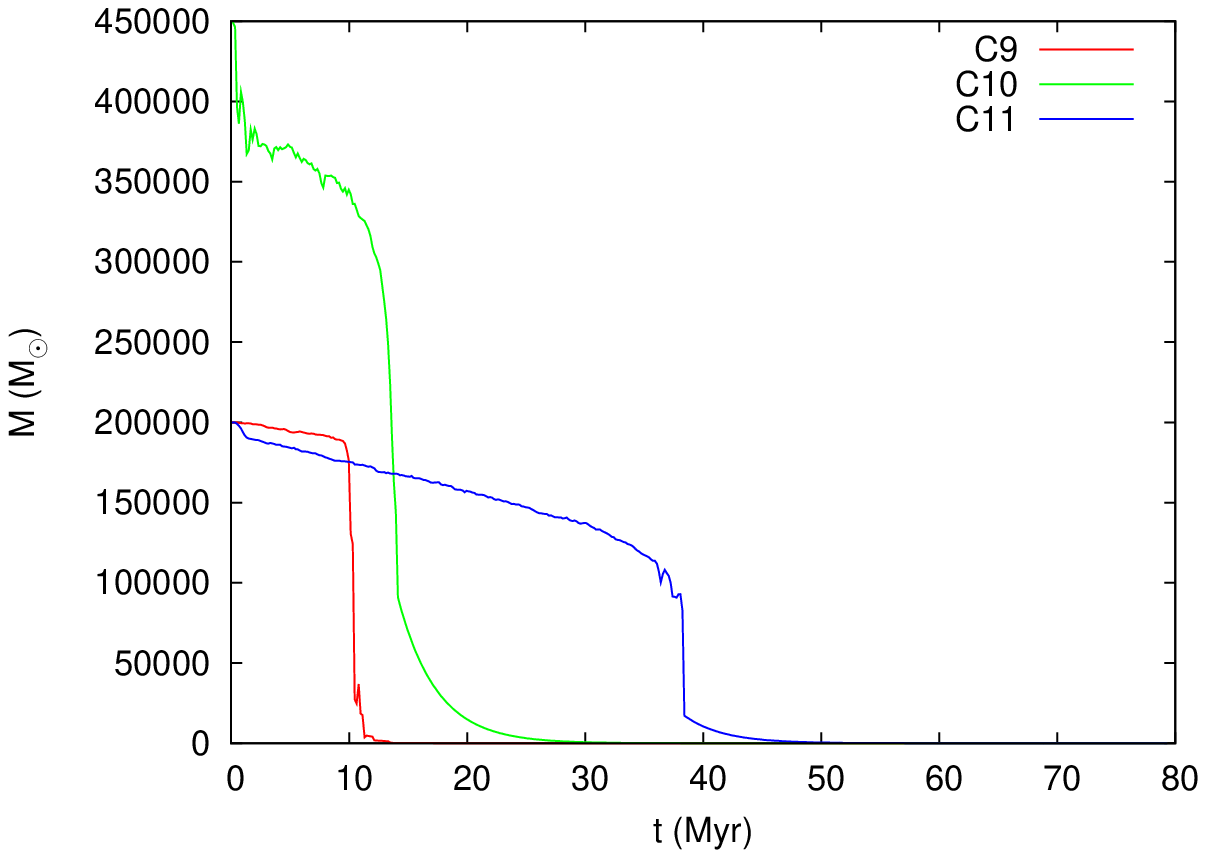}
\includegraphics[width=8cm]{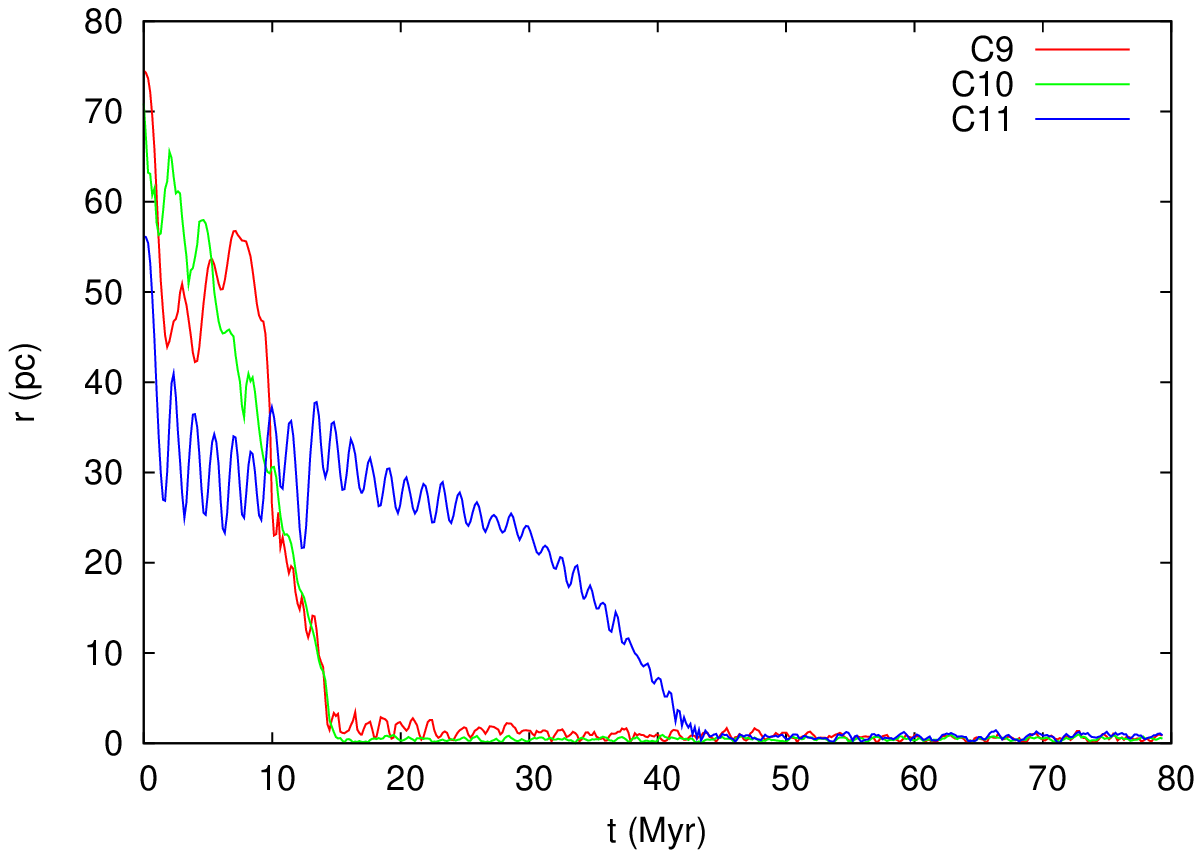}
}
\caption{As in Figure \ref{traj1}, but for simulation S2. In this case all the clusters contribute to the accretion of a nucleus.}
\label{traj2}
\end{figure*}

Moreover, clusters that reach the center transport a significant fraction of their mass toward the nucleus, as shown in the right panels of Figure \ref{traj2} and in Table \ref{tab7}. 


The distance of the BH from the galactic center as a function of time is shown in Figure \ref{BHt}. After some initial oscillations around its initial position with amplitude $\sim 2$ pc, the BH returns to the galactic center within $20$ Myr. The damping of the oscillations is due mainly to dynamical friction induced by the background stars.

\begin{figure}
\centering
\includegraphics[width=8cm]{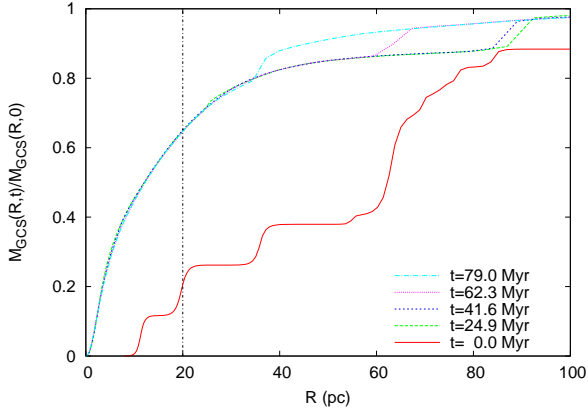}\\
\caption{Cumulative projected mass distribution of the SCS at different times in the configuration S2. The x-axis reports the projected distance from the galactic center. Here the deposited mass reach a saturation value on time-scales smaller than $20$ Myr.}
\label{mvsr2}
\end{figure}

The cumulative radial mass distribution of the SCS at different times in this S2 model is reported in Figure \ref{mvsr2}. It is interesting to note that after $\sim30$ Myr the mass distribution enclosed within $20$ pc from the BH does not change significantly since the dynamical friction timescale in this configuration does not exceed $20$ Myr for all the clusters. The only exception is cluster C11, whose decay time is $\sim 40$ Myr.\\

The mass deposited within $4$, $10$ and $20$ pc from the BH is shown in Figure \ref{NSCg2}. The Figure makes evident that, in this case, more than $50\%$ of the SCS total mass reaches the galactic center giving rise to a stellar nucleus, clearly visible out to $10$ pc ($0.2"$) in the surface density profile, as shown in Figure \ref{supd2}.

{Such a structure has a total mass $M_{\rm NSC}(r<10$ pc$)\simeq 6 \times 10^6$ M$_\odot$ and an effective radius $r_{\rm NSC} \simeq 2.63$ pc.

\begin{figure}
\centering
\includegraphics[width=8cm]{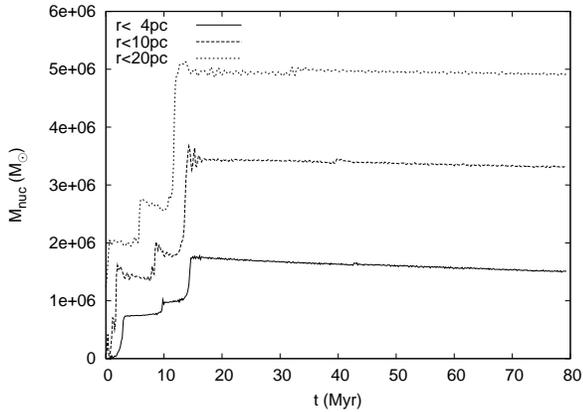}
\caption{The same as in Figure \ref{NSCg}, but for the configuration S2. }
\label{NSCg2}
\end{figure}

\begin{figure}
\centering
\includegraphics[width=8cm]{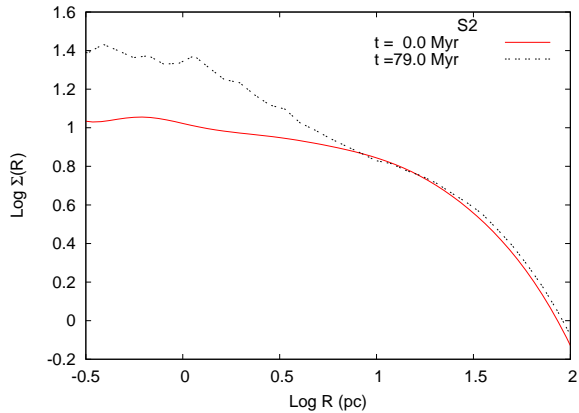}
\caption{Initial (straight line) and final (dotted line) surface density profile of the system galaxy+SCS. The x-axis reports the projected distance from the galactic center. In this case the nucleus emerges from the galactic background.}
\label{supd2}
\end{figure}

\subsection{Configuration S3}

In the simulations presented above we studied the future evolution of the Henize 2-10 central region, assuming the presence of a central BH with mass $M_{\rm BH}=2.6\times 10^6$ M$_\odot$. However, the mass measurements from \cite{reines} are based on the fundamental plane and thus quite uncertain, placing a lower limit on the mass of the BH candidate at $M_{\rm BH}=1.6 \times 10^5$ M$_\odot$.  More recent results by \cite{ngu14} give an upper limit of the BH mass of $M_{\rm BH}=10^7$ M$_\odot$.
Therefore, it is relevant understanding the role of variation of the black hole mass on the SSC evolution.

As pointed out by several authors \citep{Dolc93,antonini13,ASCD14,SASCD15}, the main effects of a central BH on the motion of infalling star clusters are that of a slight reduction of the dynamical friction decay time, enhancing at the same time the tidal disruption of the infalling clusters. 

As shown in \cite{ASCD14}, a central BH may reduce the decay time if: i) the orbit of the cluster is not far from radial, and ii) if $M_{\rm SSC}/M_{\rm BH}\simeq 1$. Otherwise, the presence of a BH does not alter significantly the clusters decay times.

On the other hand, the tidal forces induced by the BH may quench the NSC formation by cluster mergers, disrupting the incoming clusters before they reach the galactic center. Recent work has demonstrated that whenever the infalling cluster has a mass $M_{\rm SSC} <0.01M_{\rm BH}$, the tidal forces induced by the BH efficiently disrupt the cluster as it moves within the BH influence radius, thus suppressing the formation of a NSC \citep{antonini13,ASCD14b,SASCD15}. 

To investigate the quantitative role of the BH in the NSC formation process, we decided to study the extreme case of absence of a central BH in a numerical simulation where all the other initial conditions are as in model S1.

The main effect of the actual presence of a MBH can be argued from Figure \ref{traj4}, which shows the time evolution of the radial distance of SSCs from the galactic center and of the SSCs masses. A comparison of this figure with Figure \ref{traj1} shows how the MBH, on one side, affects just slightly the motion of the SSCs, but, on another side, it plays a significant role in determining the amount of mass dragged to the center of the host galaxy.
Indeed, while the time-scale for the orbital decay is almost the same in the two (S1 and S3) configurations, we found that in model S3 the SSCs reach the galactic center keeping a mass  larger than in S1. For instance, in S3 clusters C1 and C2 reach the Henize 2-10 center conserving, respectively, more than the $93\%$ and $89\%$ of their initial masses, while in simulation S1 the percentage of the bound mass is $80\%$ for cluster C1 and $\sim 80\%$ for cluster C2. This is likely due to the tidal forces induced by the MBH, which facilitate the mass erosion of  the infalling clusters. 
\\
In Table \ref{tab7} we listed the fraction of mass of each cluster left around the galactic center in configuration S3. 
Comparing such values with those of S1 makes clear that the most massive clusters, which reach earlier the galaxy center, have final masses larger than those in configuration S1 while, on the other hand, lighter clusters are almost completely disrupted before they get to the innermost region of the galaxy, with the only exception of cluster C10, which keeps $30\%$ of its initial mass by the end of the orbital decay in both the simulations.
\\
Such results can be interpreted as follows. The most massive clusters, having shorter decay time, reach rapidly the galactic center, where their evolution is determined by the combined tidal action of the galactic background and of the MBH. On the other hand, lighter clusters spend most of their time relatively far from the galactic center, where the tidal forces induced by the MBH are less efficient. Due to this, the presence of an MBH does not influence relevantly the evolution of clusters C5-C9 and C11, which are progressively disrupted by the tidal forces exerted by the galactic background. 
This is furtherly clarified by the evolution of cluster C10, which reaches the galactic center with the same final mass in both simulation S1 (where the MBH is present) and S3 (where it is not). Actually, in the configuration S1 the distance between C10 and the MBH never falls below $20$ pc, a region which encloses a stellar background mass ($\sim 1.5\times 10^7$ M$_\odot$) much larger than $M_{\rm BH}$. Hence, the evolution of clusters with long decay times, which move sufficiently far from the MBH, is mostly determined by the tidal forces induced by the stellar background, while the effect due to the MBH is negligible.

We try to resume and generalize our understanding of the effect of the presence and abscence of a MBH as studied in this paper.

In our simulations, we showed that tidal heating leads to significant disruption of clusters C5-C11 in all the cases studied.  
A BH candidate with a mass $M_{\rm BH}=10^7$ M$_\odot$ will disrupt those clusters more efficiently, but clearly this does not change the amount of mass that can be deposited toward the galactic center, since clusters C5-C11 are disrupted also by lighter BHs and only a very tiny fraction of their initial mass contributes to the growth of a bright nucleus.
Moreover, in this case the heaviest clusters, C1-C4, have masses $M_{\rm SSC}\geq 0.09 M_{\rm BH}$, and therefore tidal disruption process is not efficient enough to prevent the clusters decay, but may lead to a decrease in the accumulated matter of $10\%$ \citep{SASCD15}.

As we showed above, simulation S1 and S3 refer to the same set of initial conditions for the SCS, but S1 model contains a central BH while S3 does not. The case in which $M_{\rm BH}=1.6 \times 10^5$ M$_\odot$ lies in this two possible models. Therefore, keeping models S1 and S3 as terms of comparison, we expect that the amount of mass deposited within $20$ pc from the galactic center in such a case, should be in the range $M_{\rm dep} = 4\times 10^6$ M$_\odot$ (model S1) and $M_{\rm dep} = 4.5\times 10^6$ M$_\odot$ (model S3).

\begin{figure*}
\centering
\subfigure{
\includegraphics[width=8cm]{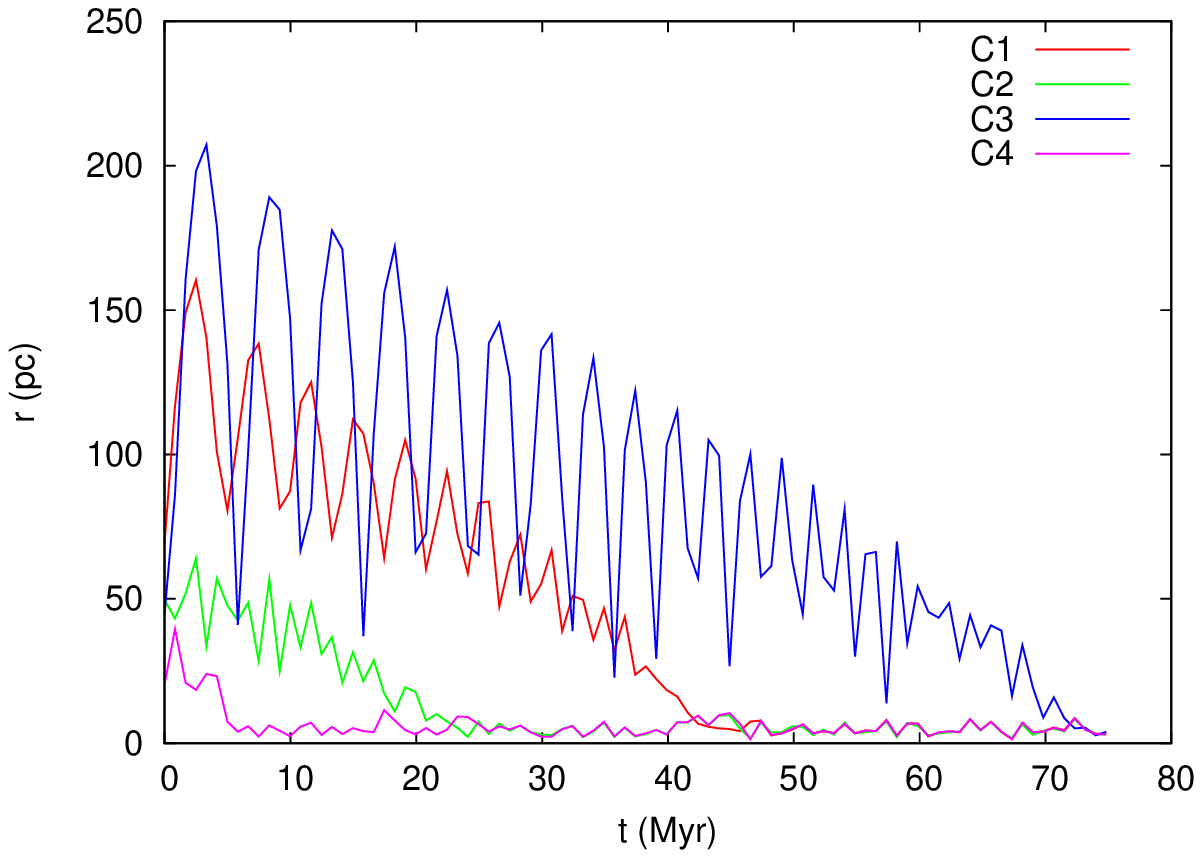}
\includegraphics[width=8cm]{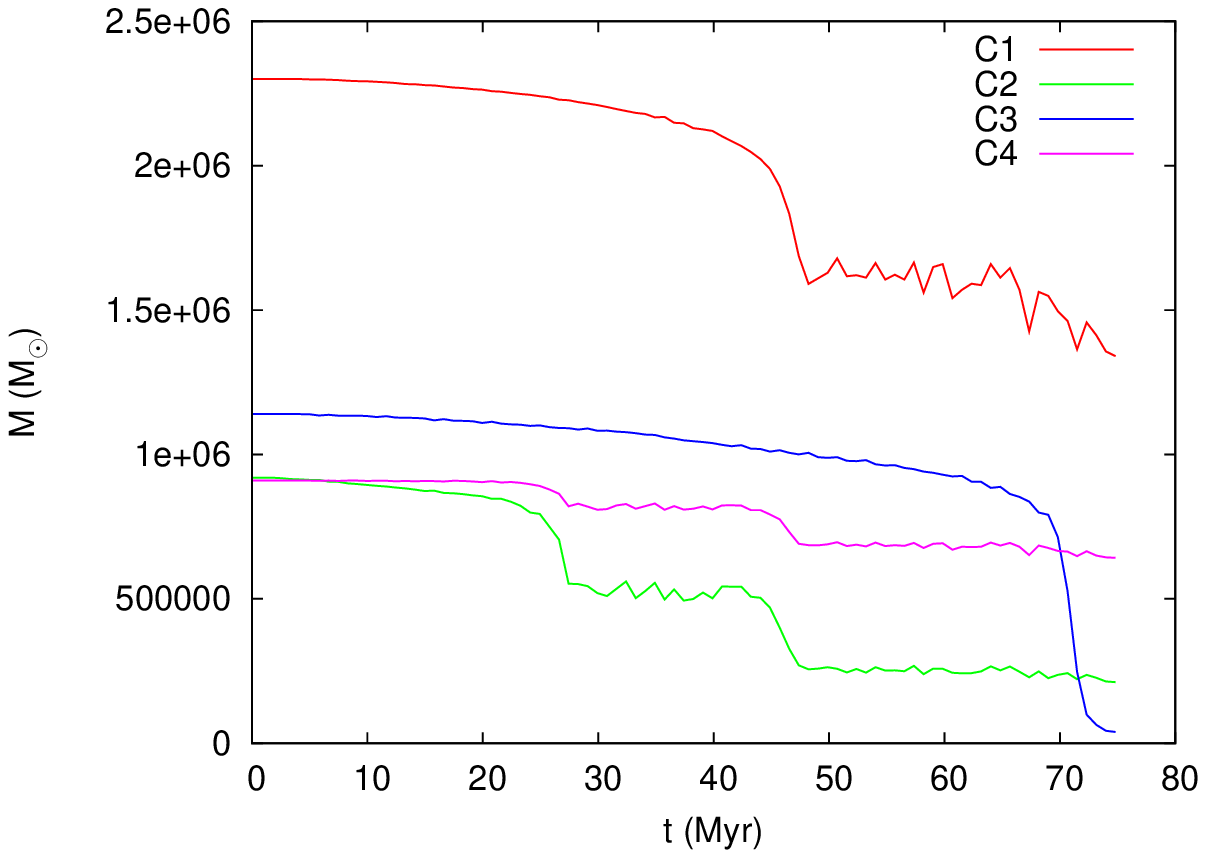}
}
\subfigure{
\includegraphics[width=8cm]{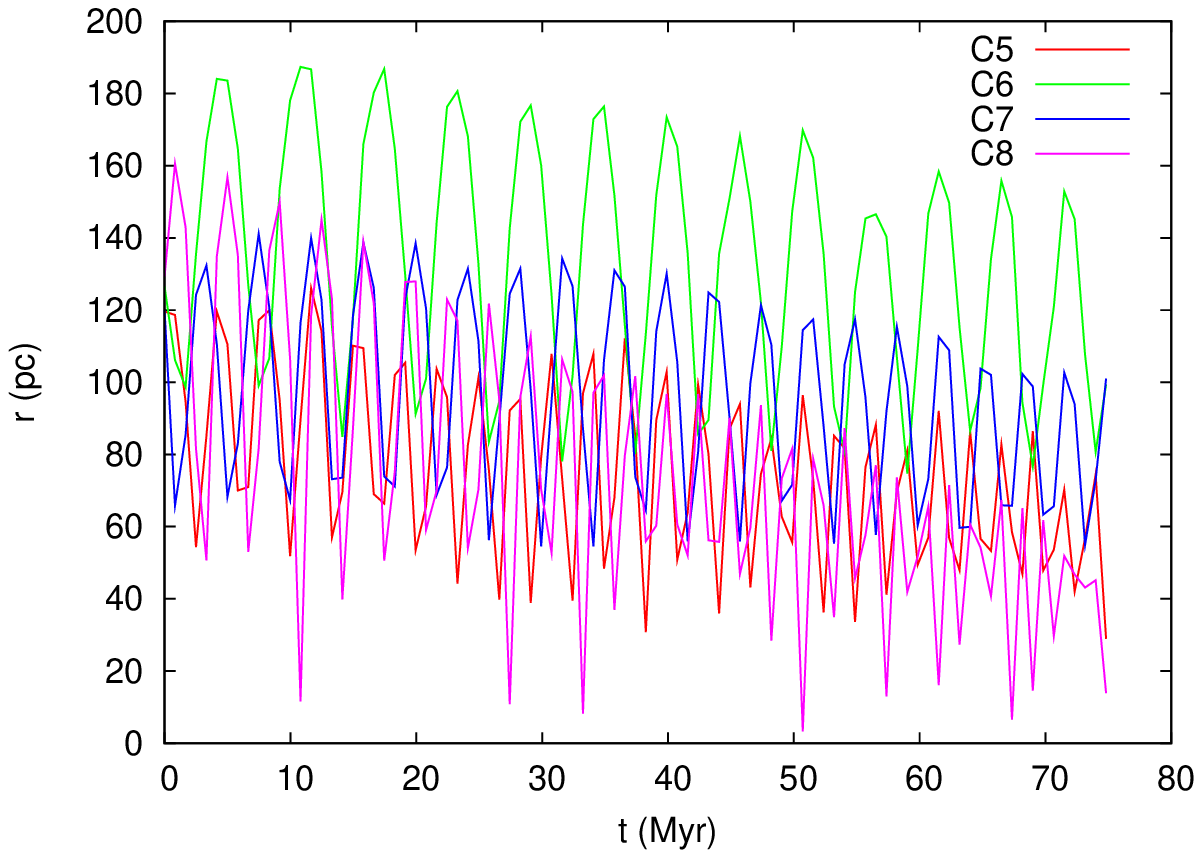}
\includegraphics[width=8cm]{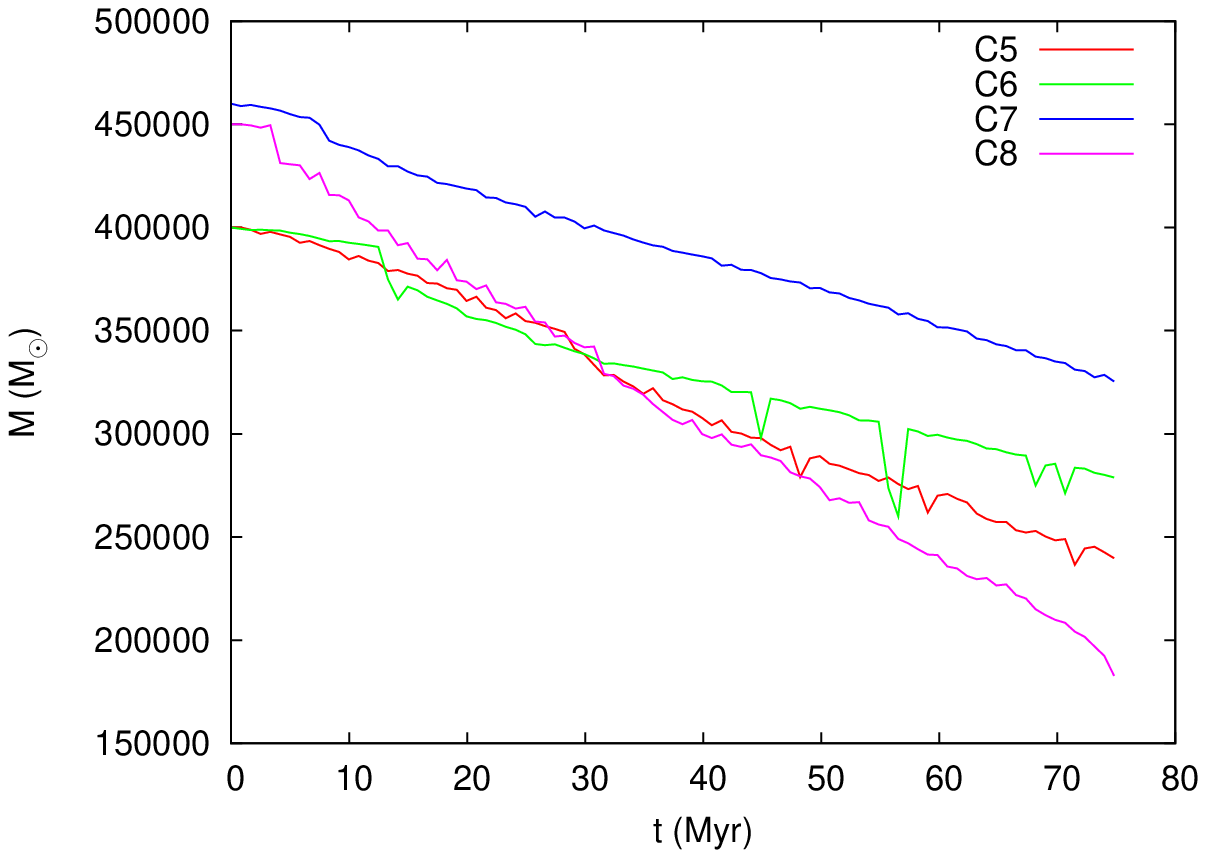}
}
\subfigure{
\includegraphics[width=8cm]{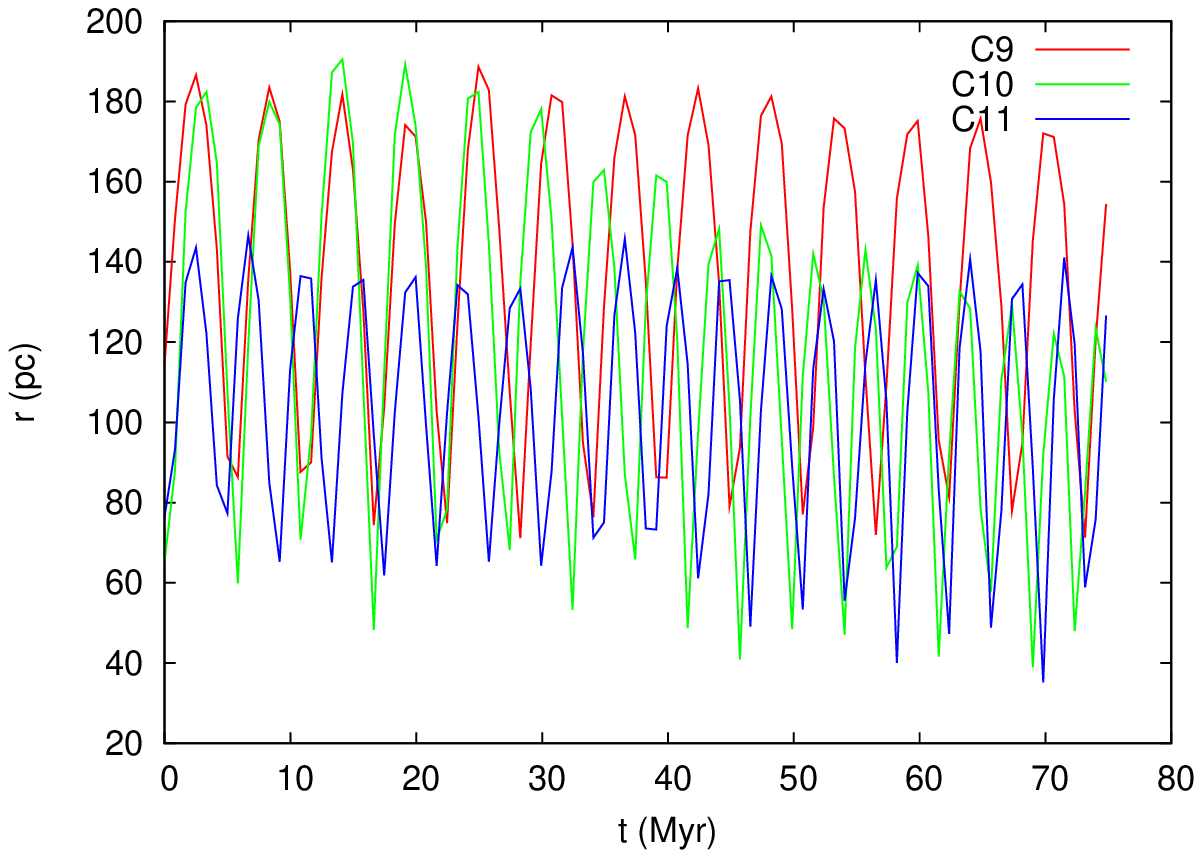}
\includegraphics[width=8cm]{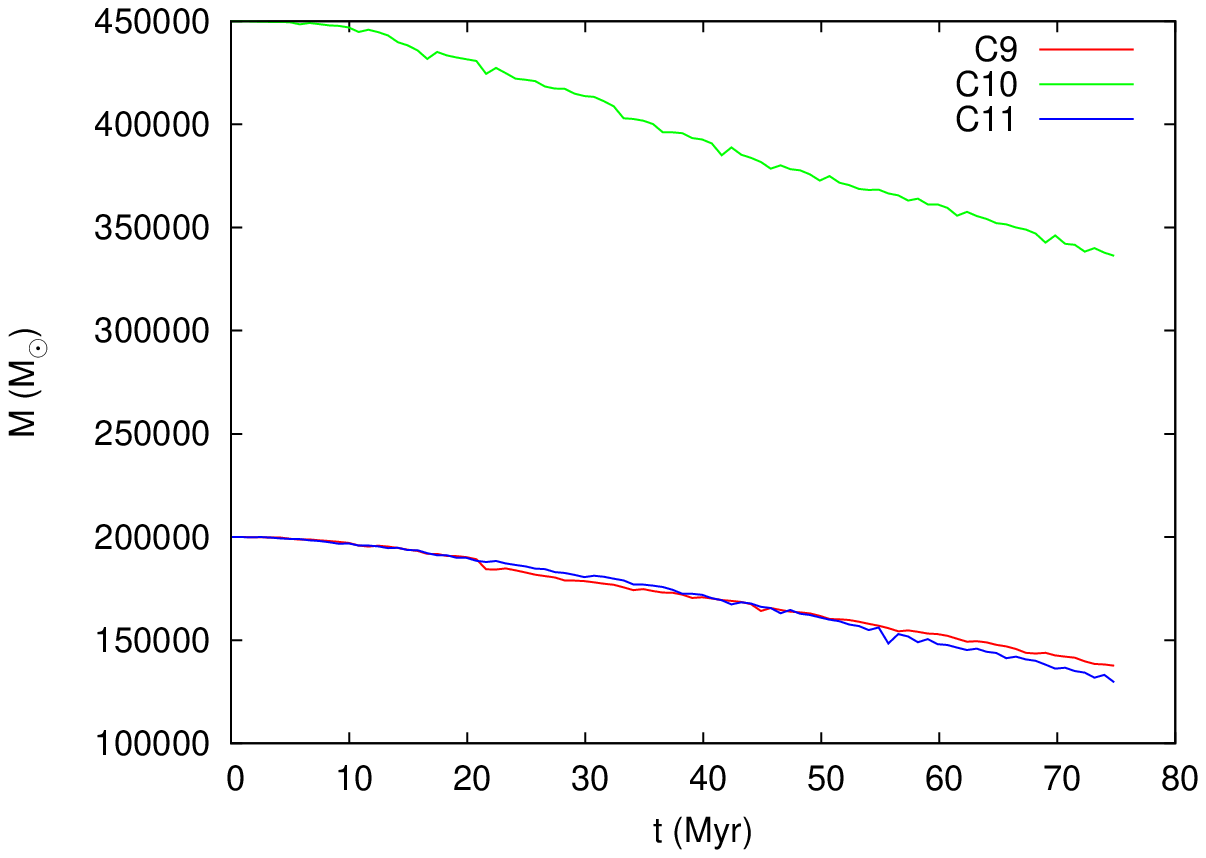}
}
\caption{As in Figure \ref{traj1}, but in the case of configuration S3. In this model the set of initial conditions for the SSCs is the same as in configuration S1, but the galaxy model does not host a MBH at its center. As consequence, the clusters reach the galactic center with a fraction of their initial mass larger than the estimates obtained from configuration S1.}
\label{traj4}
\end{figure*}


Figure \ref{mvsr4}, which shows the cumulative projected mass distribution of the SCS at different times, provides further information about the amount of mass left to the galactic center. In particular, it is evident that, over a time-scale $t\gtrsim 40$ Myr, the mass deposited within a projected radius $R = 20$ pc rises to a saturation value of $\sim 5\times 10^6$ M$_\odot$. The subsequent evolution, instead, leads to a further concentration of the SCS. Indeed, at later times, most of the deposited mass moves toward the first $10$ pc of the galaxy.

\begin{figure}
\centering
\includegraphics[width=8cm]{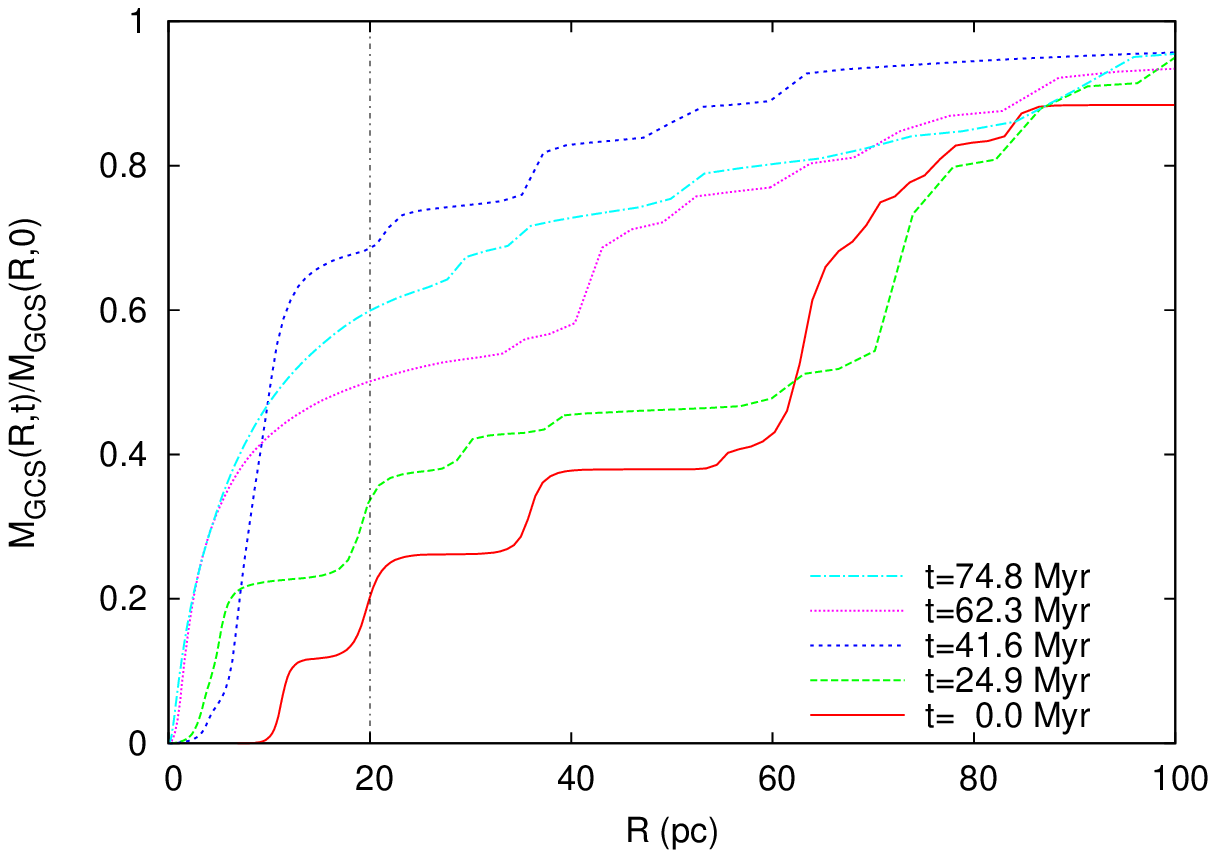}\\
\caption{Cumulative projected mass distribution of the SCS at different times in the configuration S3. The x-axis reports the projected distance from the galactic center.
Most of the deposited mass is packed within $R=20$ pc from the cluster center within $\sim 40$ Myr, while at later time such mass concentrates toward the innermost region ($R<10$ pc).}
\label{mvsr4}
\end{figure}

The mass deposited, $M_{\rm nuc}$, within $4$, $10$ and $20$ pc from the center of the galaxy is shown in Figure \ref{NSCg4}. A comparison with Figure \ref{traj4} makes clear that the growth of $M_{\rm nuc}$ is due to the decay of the most massive clusters. Indeed, each step-like increase corresponds to the decay of one of the most massive clusters. In particular, the first increase at $t\simeq 7$ Myr corresponds to the decay of cluster C4, the second to the decay of cluster C2, and, finally, the last two steps correspond to the decay of clusters C1 and C3, respectively.
Furthermore, Figure \ref{supd4} shows the initial and final surface density profile of the SCS. In this configuration, it is well evident the formation of a structure that extends up to $\sim 6$ pc ($0.16"$), enclosing a mass $M_{\rm NSC}\simeq 5.1\times 10^6$ M$_\odot$, and characterized by an effective radius $r_{\rm NSC} = 2$ pc.

\begin{figure}
\centering
\includegraphics[width=8cm]{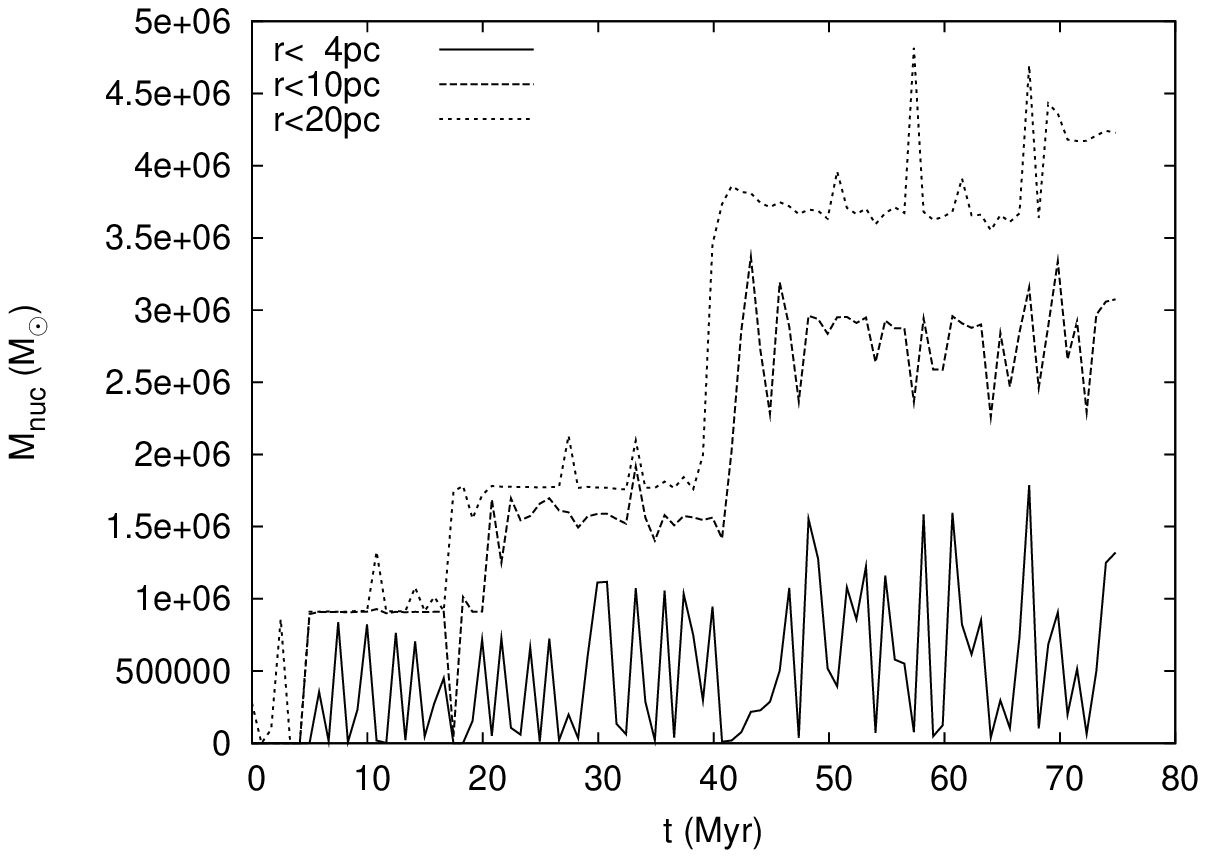}
\caption{The same as in Figure \ref{NSCg}, but for the configuration S3. }
\label{NSCg4}
\end{figure}

\begin{figure}
\centering
\includegraphics[width=8cm]{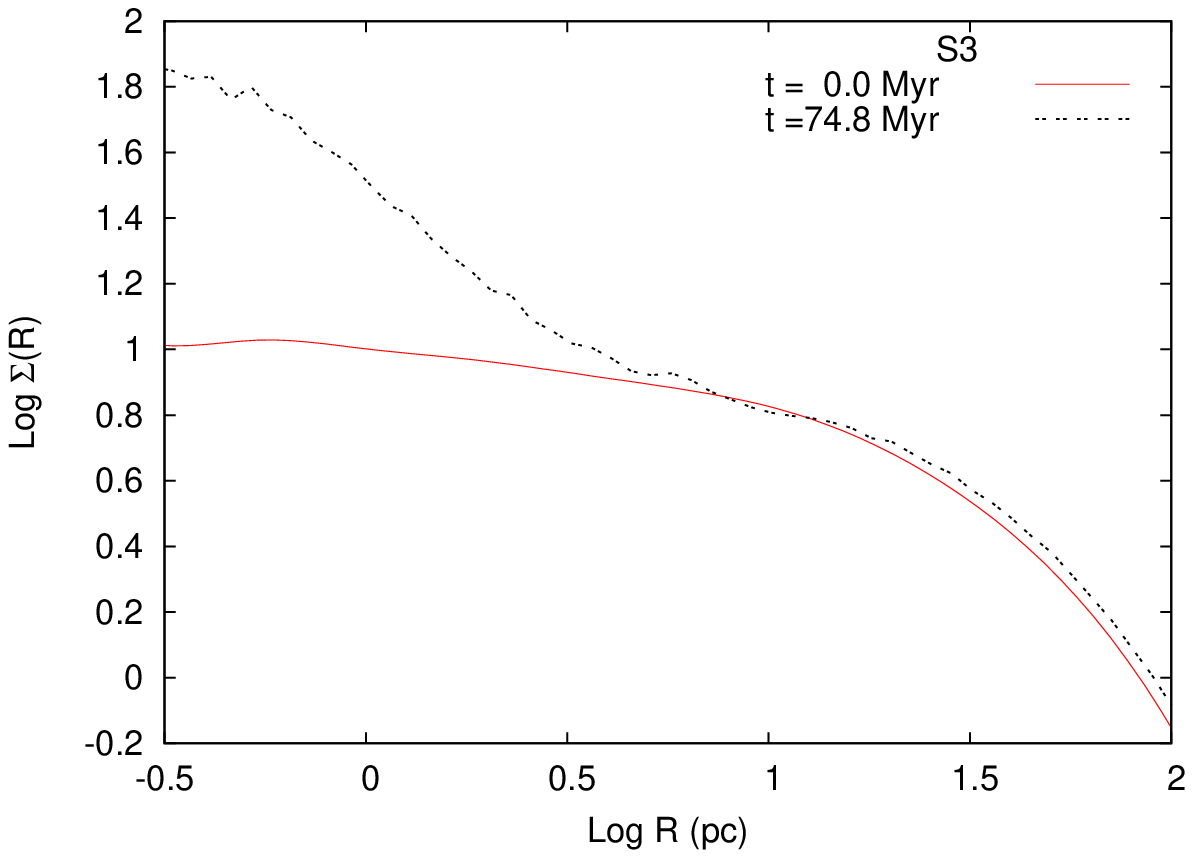}
\caption{Initial (straight line) and final (dashed line) surface density profile of the system galaxy+SCS in the simulation S3. The x-axis is the projected distance from the galactic center. It is clearly visible an overdensity that extends up to $\sim 6$ pc, with a total mass of $M_{\rm NSC}\simeq 5.1\times 10^6$ M$_\odot$ and half-light radius $r_{\rm NSC}\simeq 2$ pc.}
\label{supd4}
\end{figure}

\section{Hints on the possible long-term evolution of the nucleus}
\label{long}

Despite a detailed study of how the newly formed NSC and the surrounding nucleus will evolve and interact would require times too long to be simulated with high precision simulations, we can provide some hints about the subsequent evolution of the dense galactic center using the results obtained above.

In particular, in our $N$-body simulations we found that the mass enclosed within $20$ pc saturates to a value $\sim 4\times 10^6$ M$_\odot$ for the simulation S1, $\sim 5\times 10^6$ M$_\odot$ for the simulation S2 and $\sim 5\times 10^6$ M$_\odot$ for the simulation S3. 

Moreover, it is worth noting that the clusters which still were on their orbit at the end of the simulations, are expected to give a negligible contribution to the growth of the nucleus, since they are likely disrupted before they can reach the inner region of the galaxy (see Table \ref{tab7}).

Therefore, if we accept that such mass concentrates over a relaxation time, we can compare such estimates with the expected mass of a NSC as obtained by extrapolation of the available NSC-galaxy host observed correlations. Indeed, many authors have provided scaling laws connecting the NSC mass with the host quantities (total mass, bulge mass, velocity dispersion) using high-resolution observations \citep{BKR02,cote06,frrs,bekkiGr,Seth,LGH,Turetal12,ERWGD,scot}, or theoretical arguments \citep{LGH,antonini13,ASCD14b}. 
Using the \cite{LGH} correlation between the NSC mass and the host galaxy mass:
\begin{equation}
{\rm Log}(M_{\rm NSC}/{\rm M}_\odot)=1.18{\rm Log}(M_g/{\rm M}_\odot)-4.376,
\end{equation}

and the theoretical correlation found by \cite{ASCD14b}:
\begin{equation}
{\rm Log}(M_{\rm NSC}/{\rm M}_\odot)=1.049{\rm Log}(M_g/{\rm M}_\odot)-3.23,
\end{equation}

and assuming for Henize 2-10 $M_g = 6.4\times 10^9$ M$_\odot$, we found $M_{\rm NSC}\sim 10^7$ M$_\odot$ using both the equations above. Under the hypothesis that the mass enclosed within $20$ pc can concentrate over a relaxation time, and adding the mass of the background galaxy contained within $10$ pc to the available mass in form of star cluster, we found a central mass of $M_{\rm NSC}\sim 8 \times 10^6$ M$_\odot$, a value in good agreement with that expected from scaling relations.

\section{General remarks and conclusions}
\label{end}
The aim of this work was the investigation of the possible, future formation of a bright nucleus in the galaxy Henize 2-10 via orbital decay and merger of some of its stellar clusters. As an initial step, we used semi-analytical estimates, showing that the whole SCS of this galaxy should ``collapse'' to the galactic center in a time ranging between $\sim 0.1$ Gyr to $\sim 1$ Gyr, slightly dependent on the initial conditions chosen for the set of clusters. The predicted mass contribution of orbitally segregated clusters to the galactic nucleus has been estimated in the range $4\times 10^6$ M$_\odot$ to $6\times 10^6$ M$_\odot$ dependending on the initial conditions.

To test quantitativaley these prediction we performed direct summation $N$-body simulations. These simulations constitute a direct test of the dry-merger scenario, and show that the decay and merging of star clusters would actually give rise to a detectable, bright nucleus whose mass is compatible with the observed masses of NSCs.

The aim of the paper was also to understand whether and how the presence of a central massive BH would affect the formation process of a NSC in Henize 2-10.
The role of a BH in the galactic center has been already studied and shown to have a relevance on the infalling clusters  (see for example \cite{AMB,AntMer12,ASCD14,antonini14}). The effect consists is both a modulation of the dynamical friction on orbiting cluters and an increased efficiency of the tidal erosion process, leading (in some cases) to disruption of clusters before they reach the galactic center \citep{TrOsSp,antonini13,ASCD14b} thus reducing or even preventing the formation of a NSC therein.  
\\
In one of the simulations performed here (S1), we showed that it is possible to form a quite evident, bright, nucleus in the Henize 2-10 center within about $100$ Myr.
\\
Since observations seem to suggest that the Henize 2-10 clusters lie on the same plane, we investigated such a possibility in simulation S2. In this case, the mass deposition to the central galactic region is significantly larger than in configuration S1, due to the smaller apocenters of the SSCs orbits.
\\
Once the NSC has been formed around the galactic center, its secular evolution is mainly driven by two-body relaxation, which brings to a further central concentration of the nucleus. This might lead to the  formation of a very dense nucleus and, perhaps, to a slow accretion of stars onto the MBH. 
\\

Our main results can be summarized as follows:
\begin{itemize}
\item We created a model of the inner region of the galaxy Henize 2-10 using the highest resolution data available, representing this central region with more than $10^6$ particles, which guarantees a high level in our $N$-body modelization.
\item Our direct summation $N$-body simulations show that the future evolution of the SCS in such galaxy will likely lead to the formation of a bright nucleus on a time-scale smaller than $1$ Gyr.  The growth of the nucleus is maximized if the clusters lie, initially, on the same plane with nearly circular orbits.
\item The major contribution to the NSC formation is given by the merger of heaviest clusters (C1-C4), while lighter clusters experience a strong mass loss that leads to an almost complete disruption before they can reach the BH.
\item The decay and merging of massive clusters leads to a clearly detectable nucleus both in simulations S1 and S2, whose projected density profiles extends out to $10$ pc from the BH and whose effective radius is in the range $2-5$ pc. Intriguingly, we found that the bright nucleus has a mass comparable to the BH mass.
\item The ``future" NSC in simulations S1 and S2 will have a mass in agreement with the estimates given by observational and (approximated) theoretical estimates.
\item Motivated by the uncertainty in the evaluation of the central BH mass, we discussed the role of a variation of the mass of the BH candidate in Henize 2-10 galaxy, highlighting that, in the observed limits on the BH mass, we do not expect a significant change in the formation process of a NSC.
In configuration S3, we used the same set of initial conditions as in configuration S1 assuming, in this case, that Henize 2-10 does not host a MBH at its center. We found that the presence of a MBH slightly affects the decay process, since the full orbital decay occurs over a time which is almost the same in the two configurations, at least for the most massive clusters. On the other hand, our results showed that the MBH contributes significantly to the tidal erosion of the heaviest clusters that, in configuration S1, reach the galactic center with a mass $\sim 15\%$ smaller than in configuration S3. 
\item Finally, we found that the mass-loss process for the lighter clusters is mostly determined by the tidal forces of the galactic background. Indeed, lighter clusters are efficiently disrupted even in the configuration S3. This is likely due to the fact that clusters C5-C11 spend most of their time on orbits that never get very close to the galactic center, where the tidal effect arising from the MBH is maximized, and therefore the disruption process is mostly due to the tidal forces induced by the galactic background.
\end{itemize}

\section{Acknowledgements}
MAS acknowledges the support provided by MIUR through the grant PRIN 2010 LY5N2T 005.

\footnotesize{
\nocite{*}
\bibliographystyle{mn2e}
\bibliography{ASetal2015}
}
\end{document}